\documentclass[]{aastex631}
\usepackage{CJK}
\shorttitle{AASTeX v6.31 Sample article}
\shortauthors{Chen et al.}
\graphicspath{{./}{figures/}}
\usepackage{soul}
\usepackage{color,xcolor}
\usepackage{amsmath}
\usepackage{boondox-cal}
\usepackage{ulem}

\begin{document}
\begin{CJK*}{UTF8}{gbsn}

\title{Can the Parker Solar Probe Detect a CME-flare Current Sheet?}

\author[0000-0002-8077-094X]{Yuhao Chen}
\affiliation{Yunnan Observatories, Chinese Academy of Sciences, P.O. Box 110, Kunming, Yunnan 650216, People's Republic of China}
\affiliation{University of Chinese Academy of Sciences, Beijing 100049, People's Republic of China}

\author{Zhong Liu}
\affiliation{Yunnan Observatories, Chinese Academy of Sciences, P.O. Box 110, Kunming, Yunnan 650216, People's Republic of China}
\affiliation{University of Chinese Academy of Sciences, Beijing 100049, People's Republic of China}
\affiliation{Center for Astronomical Mega-Science, Chinese Academy of Sciences, Beijing 100012, People's Republic of China}

\author[0000-0002-7289-642X]{Pengfei Chen}
\affiliation{School of Astronomy and Space Science and Key Laboratory of Modern Astronomy and Astrophysics, Nanjing University, Nanjing, Jiangsu 210023, People's Republic of China}

\author[0000-0001-5948-559X]{David F. Webb}
\affiliation{Institute for Scientific Research, Boston College, Chestnut Hill, MA, USA}

\author[0000-0002-9264-6698]{Qi Hao}
\affiliation{School of Astronomy and Space Science and Key Laboratory of Modern Astronomy and Astrophysics, Nanjing University, Nanjing, Jiangsu 210023, People's Republic of China}

\author[0000-0001-9828-1549]{Jialiang Hu}
\affiliation{Yunnan Observatories, Chinese Academy of Sciences, P.O. Box 110, Kunming, Yunnan 650216, People's Republic of China}
\affiliation{University of Chinese Academy of Sciences, Beijing 100049, People's Republic of China}

\author[0000-0002-1264-6971]{Guanchong Cheng}
\affiliation{Yunnan Observatories, Chinese Academy of Sciences, P.O. Box 110, Kunming, Yunnan 650216, People's Republic of China}
\affiliation{University of Chinese Academy of Sciences, Beijing 100049, People's Republic of China}

\author[0000-0001-9650-1536]{Zhixing Mei}
\affiliation{Yunnan Observatories, Chinese Academy of Sciences, P.O. Box 110, Kunming, Yunnan 650216, People's Republic of China}
\affiliation{Center for Astronomical Mega-Science, Chinese Academy of Sciences, Beijing 100012, People's Republic of China}

\author[0000-0002-5983-104X]{Jing Ye}
\affiliation{Yunnan Observatories, Chinese Academy of Sciences, P.O. Box 110, Kunming, Yunnan 650216, People's Republic of China}
\affiliation{Center for Astronomical Mega-Science, Chinese Academy of Sciences, Beijing 100012, People's Republic of China}

\author{Qian Wang}
\affiliation{School of Information Science and Engineering, Yunnan University, Kunming, Yunnan 650500, People's Republic of China}

\author[0000-0002-3326-5860]{Jun Lin}
\affiliation{Yunnan Observatories, Chinese Academy of Sciences, P.O. Box 110, Kunming, Yunnan 650216, People's Republic of China}
\affiliation{University of Chinese Academy of Sciences, Beijing 100049, People's Republic of China}
\affiliation{Center for Astronomical Mega-Science, Chinese Academy of Sciences, Beijing 100012, People's Republic of China}

\correspondingauthor{Yuhao Chen}
\email{chenyuhao@ynao.ac.cn}

\received{July 2, 2023}
\revised{August 23, 2023}
\accepted{September 10, 2023}




\begin{abstract}

A current sheet (CS) is the central structure in the disrupting magnetic configuration during solar eruptions. More than 90\% of the free magnetic energy (the difference between the energy in the non-potential magnetic field and that in the potential one) stored in the coronal magnetic field beforehand is converted into heating and kinetic energy of the plasma, as well as accelerating charged particles, by magnetic reconnection occurring in the CS. However, the detailed physical properties and fine structures of the CS are still unknown since there is no relevant information obtained via in situ detections. The Parker Solar Probe (PSP) may provide us such information should it traverse a CS in the eruption. The perihelion of PSP's final orbit is located at about 10 solar radii from the center of the Sun, so it can observe the CS at a very close distance, or even traverses the CS, which provides us a unique opportunity to look into fine properties and structures of the CS, helping reveal the detailed physics of large-scale reconnection that was impossible before. We evaluate the probability that PSP can traverse a CS, and examine the orbit of a PSP-like spacecraft that has the highest probability to traverse a CS.

\end{abstract}



\section{Introduction} \label{sec:intro}

The Parker Solar Probe (PSP, \citealt{2016SSRv..204....7F}) provides us an approach to closely encountering with the Sun, and to conduct in situ measurements of various magnetic configurations/structures, plasma properties \citep{2019Natur.576..232H}, and the related phenomena in the solar atmosphere. These structures and phenomena include helmet streamers \citep{2021PhRvL.127y5101K}, magnetic fields inside the solar wind (\citealt{2019Natur.576..228K}; \citealt{2019Natur.576..237B}), energetic particles \citep{2019Natur.576..223M}, the boundary between fast and slow solar winds \citep{2023ApJ...944..116L}, the magnetized plasma bubble of the coronal mass ejection (CME, e.g., see also \citealt{2004ApJ...602..422L}, \citealt{2004NewA....9..611L}, \citealt{2011LRSP....8....1C}), as well as the large-scale magnetic reconnecting current sheet (CS) connecting the CME to the associated flare (see also \citealt{2000JGR...105.2375L} and \citealt{2002ChJAA...2..539L}). As the central structure that drives the major solar eruption, the CME-flare CS involves fundamental physics of the energy release in the most violent process in the solar system. It is very likely to include something brand new about large-scale magnetic reconnection that we do not fully understand yet. Therefore, in situ measurements of the reconnection region via traversing the CS will surely reveal the details of large-scale reconnection taking place in the CME-flare CS, and enrich the existing classical theory of magnetic reconnection and plasma physics processes throughout the universe.

\begin{figure}[ht]
\begin{center}
\includegraphics[width=0.7\textwidth]{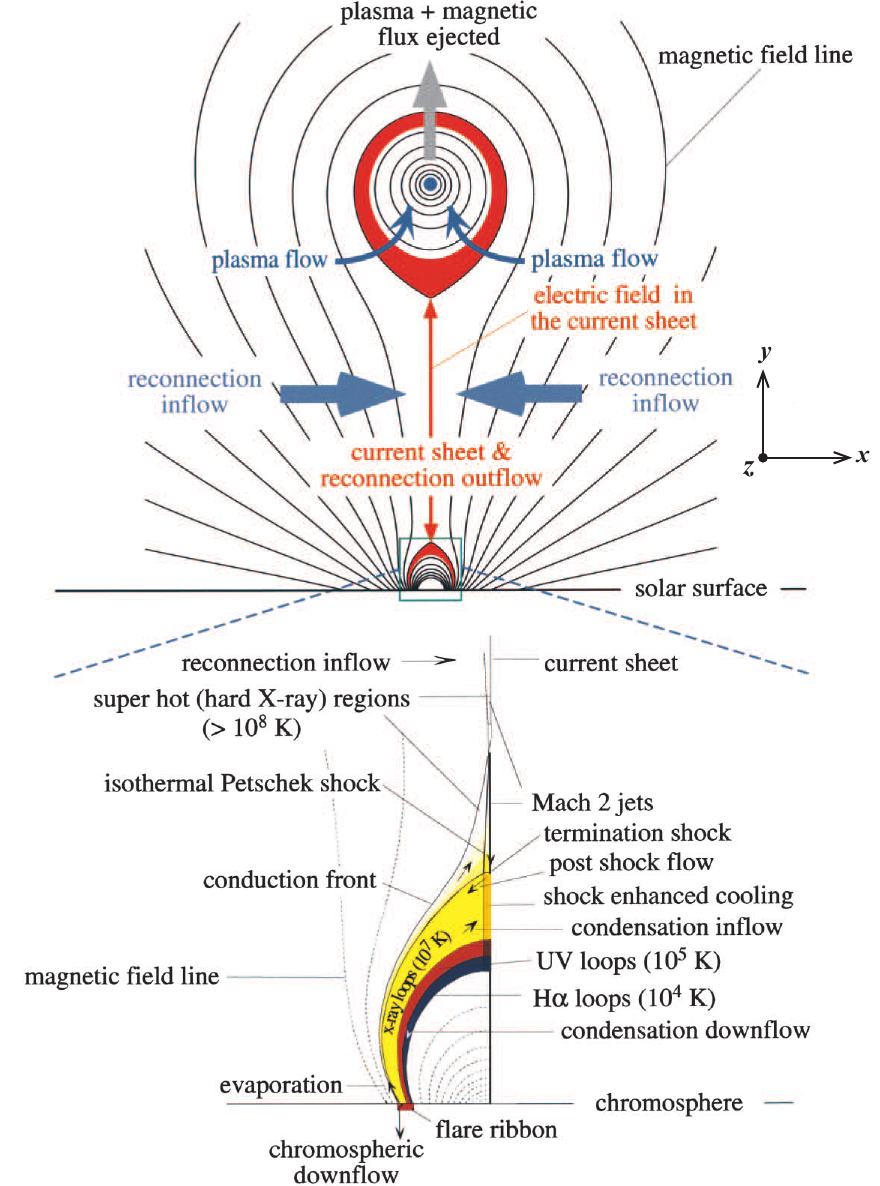}
\end{center}
\caption{A sketch of disrupted magnetic field that forms during solar eruptive process. Colors roughly indicate the plasma layers in different temperatures (from \citealt{2004ApJ...602..422L}). The diagram combines the two-ribbon flare configuration proposed by \cite{1996ApJ...459..330F}, as well as the CME configuration of \cite{2000JGR...105.2375L}.
\label{fig:CS}}
\end{figure}

Theoretically, \cite{2000JGR...105.2375L} and \cite{2002ChJAA...2..539L} pointed out that a large-scale CS forms in the eruption as the catastrophe occurs in the coronal magnetic field, the closed magnetic field is severely stretched, and two magnetic fields of opposite polarity are pushed toward each other. Magnetic reconnection then takes place inside the CS, diffuses the magnetic field, and helps the ejected magnetic structure escape from the Sun smoothly, constituting a CME. Meanwhile, magnetic reconnection also continuously produces closed magnetic field below the CS, constituting flare loops. Figure \ref{fig:CS} describes this process in an explicit fashion. Because the catastrophe occurs on the Alfv\'{e}n time scale, $\tau_{A}$, and reconnection on the diffusive time scale, $\tau_{d}$, with $\tau_{A} \ll \tau_{d}$, the CS cannot be dissipated fast enough by reconnection, and a long CS between the CME and the associated flare would be expected \citep[e.g., see also discussions of][]{2000mare.book.....P}.

As shown in Figure \ref{fig:CS}, the CS is confined to a region that is small compared to the whole disrupting configuration. This is expected since the electric conductivity is high. In the framework of the traditional theory of magnetic reconnection, the thickness of the CS must be as small as the proton Larmor radius, otherwise the fast reconnection process that is needed to drive the major solar eruptions cannot progress (e.g., \citealt{1996ApJ...462..997L,2005SoPh..226...73W}; and references therein). The proton Larmor radius is around tens of meters, not exceeding a hundred meters, in the coronal environment. After analyzing a set of unique data for several eruptions observed by Ultraviolet
Coronagraph Spectrometer (UVCS; \citealt{1995SoPh..162..313K}) and Large Angle and Spectrometric Coronagraph Experiment
(LASCO; \citealt{1995SoPh..162..357B}) on the Solar and Heliospheric Observatory (SOHO), on the other hand, \cite{2007ApJ...658L.123L,2009ApJ...693.1666L} found that, in some circumstances, the CSs are observable, and their thickness in real events could be as large as a few $10^{4}$~km or even $10^{5}$~km. Many follow-ups on this topic by different authors for different events observed in different wavelengths by different instruments both in space and on the ground gave the similar results such that the apparent thickness of the CME-flare CS ranges from $10^{3}$ to $10^{5}$~km (see \citealt{2013ApJ...766...65C,2015SSRv..194..237L}; and \citealt{doi:https://doi.org/10.1002/9781119324522.ch15} for more details). \cite{2008ApJ...686.1372C} noticed that observational data in different wavelengths for the same event gave the same value of the CS thickness. Significant difference apparently exists between the value of the CS thickness expected according to the classical theory of magnetic reconnection and that deduced from observations. Although the values of the CS thickness given by observations span two orders of magnitudes, the difference among these values is still small compared to that between several tens of meters and a few 10$^{4}$~km.

Usually, it is believed that the difference between the apparent thickness, $d^{\prime}$, of the CS and the true thickness, $d$, results from three issues: the projection effects, the complex structure of the CS, and the thermal halo around the CS. The projection effects exist for all the images we obtained via the remote sensing approach, since any image we obtained is the projection of the emission from the optically thin three-dimensional object onto the two-dimensional plane of the sky. The intensity of the emission reaching the detector is the sum of all the emission from the object in the line-of-sight (LOS), and the level of intensity is governed by both the density and the column depth in LOS. Thus, a bright object manifests differently when being seen at different angle (see detailed discussions of \citealt{1996ApJ...459..330F}). This yields that $ d^{\prime}\ge d$, and that the emission measure of the CS reaches maximum and $d^{\prime} = d$ as the CS is observed edge-on as shown in Figure \ref{fig:CS}. In principle, $d^{\prime}$ could be very large when the CS is observed face-on as shown by \cite{2006ApJ...638.1110B}. The terms ``edge-on'' and ``face-on'' here refer to two distinct observational angles: ``edge-on'' implies that the LOS is parallel to the surface of the CS, namely along the $z$-direction (see Figure \ref{fig:CS}), while ``face-on'' means that the LOS is perpendicular to the surface of the CS, namely along the $x$-direction (see Figure \ref{fig:CS}). For more information, interested readers refer to \cite{2006ApJ...638.1110B}, \cite{2007ApJ...658L.123L,2009ApJ...693.1666L}, as well as \cite{2022NatAs...6..317S}.

On the other hand, \cite{2009ApJ...693.1666L} pointed out that the emission measure of the CS is usually small compared to the nearby structures like CMEs, helmet streamers, and so on. If the viewpoint toward the sheet deviates largely from that edge-on, the CS would become too faint to be seen. They found that the emission measure of the CS is roughly related to $d/d^{\prime}$ linearly, which suggests that the projection effects on measuring $d$ are limited. Furthermore, the limited signal-to-noise ratio of the instrument also enhances the difficulty in identifying the CS in reality. Therefore, \cite{2009ApJ...693.1666L} concluded that the CS would become invisible if $d/d^{\prime} < 0.1$, and \cite{2008ApJ...686.1372C} realized that $d/d^{\prime}$ ranged from 0.2 to 0.4 for the CS developed in a specific event.

The fact that the CS may possess complex structure could also increase the value of $d^{\prime}/d$. \cite{2009A&A...499..905V} studied three events occurring on 26 June 2005, 8 January 2002, and 10 November 2003, respectively, and found that the values of $d^{\prime}$ varies from $7\times 10^{4}$~km to $2.1\times 10^{5}$. They showed that a CS forms as the associated closed coronal magnetic field is severely stretched by the eruption, the CS is thus typically highly planar, and no obvious warping occurs in the eruption although various small scale structures exist inside the sheet (see also discussions of \citealt{doi:https://doi.org/10.1002/9781119324522.ch15}). The results of \cite{2009A&A...499..905V} suggested that the impact of the complex structure of the CS on measuring $d$ may yield that $d^{\prime}$ differs from $d$ by only a factor of single digit.

The thermal halo also plays a role in broadening the CS observed in spectral lines forming at high temperature like [Fe XVIII] and [Ca XIV]. \cite{1998ApJ...494L.113Y,2001ApJ...549.1160Y} noticed the occurrence of the thermal halo for the first time such that the plasma heated by reconnection inside the CS may leak to the inflow region, constituting a thermal halo around the CS. Numerical simulations by \cite{2009ApJ...701..348S}, \cite{2010ApJ...721.1547R}, and \cite{2012ApJ...758...20N} confirmed this result, and the CS is in fact embedded in the thermal halo. This implies that $d^{\prime}$ is actually the scale of the halo, not that of the CS itself (see the detailed discussions by \citealt{2015SSRv..194..237L} and \citealt{doi:https://doi.org/10.1002/9781119324522.ch15}).

However, both observations and theories indicated that the thermal halo does not always occur in reality. The CS developed in an event studied by \cite{2008ApJ...686.1372C} was observed in both white-light and high temperature spectral lines obtained by the UVCS, and \cite{2008ApJ...686.1372C} found that the values of $d^{\prime}$ deduced from both the white-light and the high temperature spectral data are the same. Since the white-light emission of the observed target results from the scattering of the photospheric emission by the free electrons, the thermal property of the target does not affect its manifestation in white-light. This implies very limited impact of the thermal halo on measuring $d$. \cite{2017ApJ...835..139S} pointed out that the region of the thermal halo must be tenuous compared to that of the CS if the thermal halo does occur but cannot be recognized. However, \cite{2017ApJ...843..121R} argued that the hot plasma inside the CS is prohibited from leaking outside by the electric field in the slow mode shock should the Petschek reconnection take place through the CS, therefore the role of the thermal halo is often overestimated. Numerical calculations by \cite{2016ApJ...832..195N} also indicated the limited impact of the thermal halo on measuring $d$ (see also detailed discussions of \citealt{doi:https://doi.org/10.1002/9781119324522.ch15}).

The above discussions and results indicate that the reconnecting CS occurring in the solar eruption may indeed possess huge thickness, and that the projection effects, the complex structure, and the thermal halo are not able to account for difference in the CS thickness between the expectation of the classical theory of reconnection and the observational results. \cite{2015SSRv..194..237L} and \cite{doi:https://doi.org/10.1002/9781119324522.ch15} concluded that the three issues below may account for the huge thickness of the CME-flare CS. First of all, the CME-flare CS develops in a highly dynamic fashion in the solar eruption, instead of staying static. Both theories \citep{2000JASTP..62.1499F,2000JGR...105.2375L,2002ChJAA...2..539L,doi:10.1063/1.1563668} and observations \citep{2003JGRA..108.1440W,2016SoPh..291.3725W} showed that the length of the CS increases at a speed up to a few hundred km~s$^{-1}$ and at an acceleration of a few m~s$^{-2}$. Such a highly dynamic process in the large-scale magnetic configuration could impossibly be governed by individual particles.

Second, large-scale CSs suffer from several MHD instabilities, such as the tearing mode, giving rise to turbulence and a large number of small structures in the CS \citep{2011ApJ...737...14S,2012MNRAS.425.2824M,2012ApJ...758...20N,2019MNRAS.482..588Y}. These small structures enhance the diffusion of the magnetic field through the CS equivalent to adding an extra resistivity in the reconnection region, which is also known as the ``hyper\textendash resistivity'' (e.g., see also \citealt{1988ApJ...326..412S} and \citealt{1995ApJ...449..739B} for more details). In addition to the small scale structure, the large-scale CS has enough space to allow different types of reconnection to occur simultaneously, which never happens in the scenario of the traditional reconnection theory \citep{2012MNRAS.425.2824M,2017A&A...604L...7M,2021ApJ...909...45Y,2022RAA....22h5010Z}. This reminds us of the parallel computation usually used in modern numerical calculations, through which a large and complicated computing mission is divided into many small and simple ones that could be solved simultaneously in a shorter time.

Third, coalescence or merging of small-scale structures frequently occurs inside the CS as well (\citealt{2011ApJ...737...14S}; \citealt{2012MNRAS.425.2824M}; \citealt{2017ApJ...848..102T}; \citealt{2019MNRAS.482..588Y}), which is not a simple merging process but yields secondary reconnection or diffusion among these structures. The coalescence can be considered as the inverse cascading in which small scale structures merge into larger ones. In reality, the coalescence and the cascading processes take place simultaneously in the CS, and eventually reach a dynamic balance (see discussions of \citealt{2011ApJ...730...47B,2011ApJ...737...24B}). \cite{doi:10.1063/1.4816711} studied the two processes in the large-scale CS, and found that the kinetic energy of the plasma flow manifests similar cascading behavior, implying the dissipation of the kinetic energy of the fluid motion.

The above discussions draw a scenario of the reconnection process taking place in the solar eruption such that the large-scale CS is an assembly of many diffusive structures that allow reconnection to occur at many places in several ways simultaneously. We realize the analogy of this process to the parallel computing that is frequently used in modern numerical calculations for complicated mathematical problems. In the parallel computing process, large and difficult calculations are divided into many small and simple ones that could be solved easily and quickly. In principle, we have so far reached the theoretical explanation why magnetic reconnection in a large-scale CS could progress fast enough to drive the major eruption. But there is no existing in situ information about the physical property and the internal structures of the CS, and the explanation cannot be finalized. It was impossible to close this logical loop before PSP. However, difficulty still exists if PSP does not traverse a CS. So it is worth investigating whether PSP could traverse a CS in the solar eruption, and what the probability of traverse in a certain fashion would be. In fact, there are already many crossings of the heliospheric CS (e.g., \citealt{2021PhRvL.127y5101K,2023ApJ...948...24L}) and CS in the magnetosphere of Earth (e.g., \citealt{2023JGRA..12831066S}) and Jupiter (e.g., \citealt{2023GeoRL..5004123X}). However, the confirmed crossing of a CME-flare CS that undergoes rapid and complex magnetic reconnection has not been reported yet. A possible crossing was reported recently by \cite{Romeo_2023} and \cite{patel2023closest} who investigated a fast CME event on September 5, 2022 that swept PSP.

Because the orbit of PSP is fixed, we cannot guarantee that PSP could possess the probability as high as expected to traverse the CS. In this work, according to the distribution of the filament and the filament channel, the orientation of the filament axis relative to the latitude line on the solar surface, the speed increase of the CS length, and the lifetime of the CS, we figure out the probability for a PSP-like spacecraft to traverse a CS \citep{2021ChJSS..41..183L} with a given orbit, and look for the orbit that would yield highest probability of traversing. In Section \ref{sec:Methods}, we describe the model, with reasonable assumptions, describing the spacecraft orbit and the orientation of CSs behind CMEs. In Section \ref{sec:Result interact}, the fashion in which the orbit of spacecraft would intersect the CS will be discussed for various types of CSs. In Section \ref{sec:Result traverse}, the probability of traversing a given CS for the spacecraft on a given orbit will be evaluated, and the orbit of the spacecraft that leads to the highest traversing probability will be further suggested. Finally, we shall discuss our results and draw our conclusions in Section \ref{sec:Discussion}.

\section{Methods} \label{sec:Methods}

In this section, we describe the mathematical approach to calculating the probability of a spacecraft (including PSP) traversing the CME-flare CS. First of all, the calculation employs two sets of coordinate systems on the basis of the ecliptic plane and the plane where the spacecraft orbit is located, respectively. Second, according to the Lin-Forbes model and the related observations, we constructed a simplified model for the CS geometry. Then, we relate the parameters of the spacecraft orbit to those of the CS; and finally, we are able to estimate the probability that a spacecraft on a given orbit traverses a CME-flare CS.

\subsection{Descriptions of Coordinate Systems} \label{sec:Coordinate}

According to the purpose of the mission, the heliocentric orbit of the existing spacecraft for solar observations and/or detections falls into two categories: one like those of Ulysses \citep[][]{1992A&AS...92..207W} and Solar Orbiter \citep[][]{2020A&A...642A...1M}, obviously deviating from the ecliptic plane, and another one like those of STEREO \citep[][]{2008SSRv..136....5K} and PSP, slightly deviating from the ecliptic plane. PSP is the first spacecraft to fly into the solar corona in human history and is, therefore, very likely to traverse CME-flare CSs. With more and more spacecraft being launched for solar exploration, it is necessary to analyze the impact of orbital parameters on realizing the spacecraft traversing and then detecting CSs, so that more scientific goals for the PSP-like missions in the future could be figured out. Therefore, this work is to evaluate the probability of PSP traversing a CME-flare CS, and to look into how a PSP-like spacecraft could traverse the CS with a reasonably high probability.

\begin{figure}[ht]
\begin{center}
\includegraphics[width=0.9\textwidth]{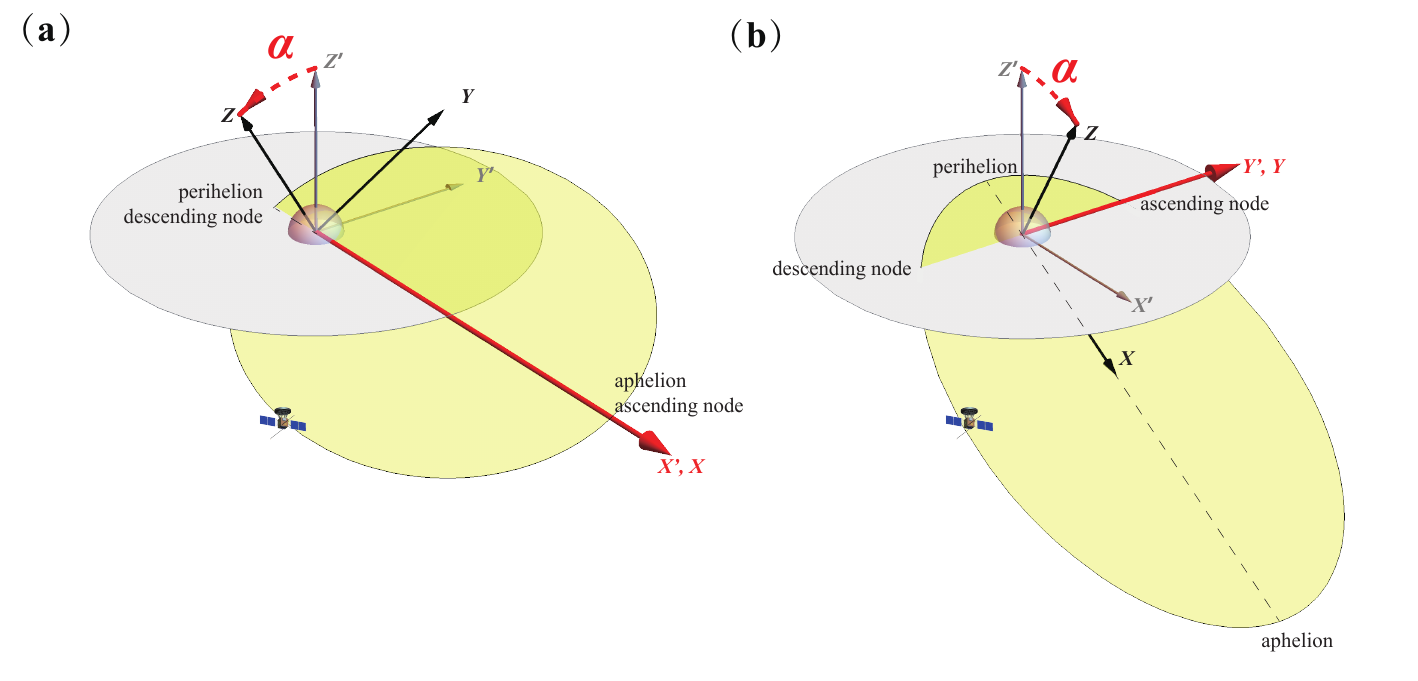}
\end{center}
\caption{The schematic diagram of two ways that the orbital plane deviates from the ecliptic plane. The gray plane is the ecliptic plane, whereas the yellow plane depicts the orbital plane. (a) The orbital plane can be obtained by rotating the inclination angle $\alpha$ around $X'$-axis, which denotes the major axis of the orbit. (b) The corresponding rotation axis is the $Y'$-axis, which represents the minor axis of the orbit.
\label{fig:Coordinate}}
\end{figure}

We set up two heliocentric coordinate systems, denoted as the ``solar coordinate system'', $X'Y'Z'$, and the ``orbital coordinate system'', $XYZ$, respectively. In both systems, the Sun is located at the origin and at one of the foci of the elliptical orbit (see Figure \ref{fig:Coordinate}), while the $X^{\prime}Y^{\prime}$-plane and $XY$-plane are coincident with the ecliptic and the orbital planes, respectively. In Figure \ref{fig:Coordinate}, the gray ellipse is the ecliptic plane, $Z'$-axis points to north and $X'$- and $Y'$-axes point toward longitudes of $\phi_s=0^\circ$ and $\phi_s=90^\circ$, respectively. On the other hand, the light yellow ellipse is the orbital plane, the $Z$-axis is perpendicular to this plane, while the $X$-axis is along the major axis of the orbit. We assume that the spacecraft moves in a counterclockwise direction around the $Z$-axis, i.e., the angular momentum of the spacecraft is parallel to the $Z$-axis. PSP moves in the plane of the Venus orbit, which slightly deviates from the ecliptic plane. We will show later that the PSP orbit is not the one that allows it to traverse the CS with a reasonably high probability. To look for the orbit with the highest probability of crossing a CME-flare CS, we introduce a parameter, $\alpha$, the angle between the orbital and ecliptic planes (see Figure \ref{fig:Coordinate}).

Figure \ref{fig:Coordinate} describes two ways by which two planes intersect with an angle of $\alpha$. The first one, which is depicted in Figure \ref{fig:Coordinate}a, involves rotating the orbital plane counterclockwise around the $X^{\prime}$-axis by an angle $\alpha$, while the perihelion, aphelion, and the major axis of the orbit keep staying in the ecliptic plane and are all located on the $X^{\prime}$-axis. The perihelion and aphelion are co-located in space with the descending and the ascending nodes of the orbit at longitudes of $180^\circ$ and $0^\circ$, respectively. The second approach, illustrated in Figure \ref{fig:Coordinate}b, rotates the orbital plane around the $Y^{\prime}$-axis by an angle $\alpha$, while the $Y^{\prime}$-axis is parallel to the minor axis of the orbit ellipse. Here, the perihelion and aphelion deviate from the ecliptic plane and are located at the northernmost and southernmost points of the orbit, with the descending and the ascending nodes at longitudes of $90^\circ$ and $270^\circ$, respectively. In both cases, the angle $\alpha$ represents the orbital inclination.

To provide a quantitative description of the two orbits shown in Figure \ref{fig:Coordinate}, we describe them mathematically in the $XYZ$ system first such that the orbit ellipse with one focus located at the origin of the coordinate system is given below:
\begin{eqnarray}
\frac{(x-c)^2}{a^2}+\frac{y^2}{b^2}=1\label{Orbit1},\\
z=0\label{Orbit2},
\end{eqnarray}
where $a$ and $b$ are the lengths of the semi-major axis and the semi-minor axis, respectively, and $c=\sqrt{a^2-b^2}$.

In principle, a more intuitive approach to the calculation would be to take the $X'Y'$ plane as the referential plane (see Figure \ref{fig:Coordinate}), to transform the orbital Equations (\ref{Orbit1}) and (\ref{Orbit2}) from the $XYZ$ to the $X'Y'Z'$ system, and then to determine whether the orbit intersects the CS or not in the $X'Y'Z'$ system. However, this approach is obviously difficult due to the nonlinear property of Equation (\ref{Orbit1}). But it should be much easier to perform transformation inversely by rotating the CS from the $X^{\prime}Y^{\prime}Z^{\prime}$ system to the $XYZ$ system since a linear function can be used to describe a CS.

\subsection{Morphology of a CME-Flare CS} \label{sec:Current sheet}

As mentioned earlier, the global behavior of the CS is relatively simple although the internal structure of the CS could be fairly complex. This is because the CS forms as a result of the severe stretch of the closed coronal magnetic field in the initial stage of the eruption. On one hand, it elongates apparently in the direction of the CME propagation. In another two orthogonal directions, on the other hand, its scale is either confined by the reconnection inflow to a local region, governing the CS thickness, or confined by the two legs of the associated CME to a fan-like region of the finite angular width (see the discussions of \citealt{2015SSRv..194..237L}). \cite{2011LRSP....8....1C} suggested that the upper limit of the angular width of this fan-like region to be approximately $70^ {\circ}$. The shape of the CS will be simplified into a triangle-like sheet with a thickness of $d$ and an extending angle less than $70^{\circ}$. For simplicity, we consider the half angular width of a typical CS to be $23^{\circ}$. Although the selection of this value is somewhat arbitrary, it remains reasonable. In addition, we further assume that: (1) the CME erupts radially (although a small part of CMEs are ejected non-radially), (2) the CS trails behind the CME, and (3) the morphological evolutions of the CME and the CS are self-similar expansion.

\begin{figure}[ht]
\begin{center}
\includegraphics[width=1.0\textwidth]{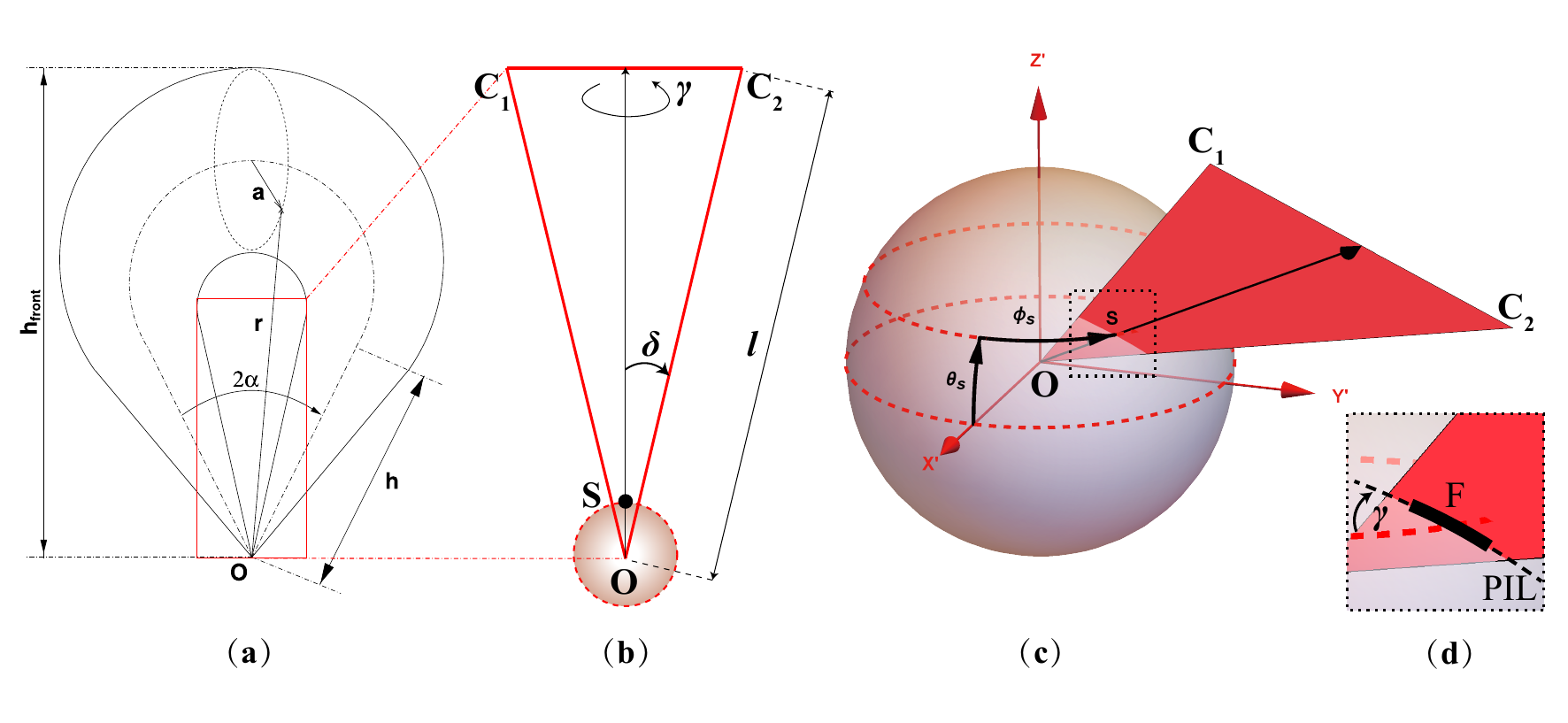}
\end{center}
\caption{The figure depicts the simplified morphology of a CS. (a) The GCS model by \cite{2009SoPh..256..111T}, where the red box highlights the location of CS. (b) The basic features of CS, including its length ($l$), half angular width ($\delta$), and tilt angle ($\gamma$). (c) The position, longitude ($\phi_s$), and latitude ($\theta_s$) of the eruption source labeled as ``S''. (d) A magnified view of the region enclosed by the dotted box in (c), where "F" denotes the filament along the northeast-southwest direction ($\gamma<0$), and the dashed line represents the polarity inversion line (PIL).
\label{fig:ModelSketch}}
\end{figure}

On the basis of these assumptions, we use the GCS model (see Figure \ref{fig:ModelSketch}a) developed by \cite{2009SoPh..256..111T} for reconstructing the CME/ICME to describe the CS morphology. As illustrated in Figure \ref{fig:ModelSketch}b, we co-locate one vertex of the triangle-shaped CS with the origin of the GCS model, which is the center of the Sun denoted as $O$. The symmetry axis of the CS intersects the solar surface at the center of the eruption source region, denoted as $S$, and two boundaries of the CS extend along $OC_{1}$ and $OC_{2}$; $\delta$ is the half angular width and $\gamma$ is the tilt angle between the local latitude and the intersecting line of the CS with the solar surface. Considering the fact that a CME usually originates from a filament or filament channel, we assume that $\gamma$ is identified with the tilt angle of the filament prior to eruption (e.g., \citealt{2011ApJ...732L..25C}, \citealt{2012NatCo...3..747Z}, and references therein). We define $\gamma>0$ if the spine of filament is along the northwest-southeast direction, and $\gamma<0$ if the filament follows the northeast-southwest direction. It should be noted that CMEs sometimes rotate as they lift off (\citealt{2023SoPh..298...35Z} for instance), so there is some uncertainty about the orientation of the CS plane in the morphology we described here. Figure \ref{fig:ModelSketch}(c) shows the relative positions of the CS and the Sun in the heliocentric coordinate system, where the longitude and latitude of point $S$, are $\phi_s$ and $\theta_s$, respectively. Figure \ref{fig:ModelSketch}d shows the enlarged area marked by the dashed box in Figure \ref{fig:ModelSketch}c, in which the black dashed line is the intersecting line of the CS with the solar surface, also known as the polarity inversion line (PIL) on the photospheric surface, and ``F'' is the location of the filament before eruption. \cite{2015ApJS..221...33H} reported that most filaments in the north hemisphere are along the northeast-southwest direction, so we let $\gamma<0$ for the filament in Figure \ref{fig:ModelSketch}d to match this rule. In addition to these parameters, those, such as the CS life-time, $\tau_{0}$, extending velocity of the CS, $v$, and the eruption initiation time, $t$, related to the dynamical properties of the eruption need to be taken into account as well.

It is necessary to determine the coordinates of $C_1$ and $C_2$ in the $XYZ$ system before the criterion is established for traversing of the spacecraft on a given orbit with a CME-flare CS. For a CS described by its spatial and morphological parameters, denoted as ($l$, $\delta$, $\phi_{s}$, $\theta_{s}$, $\gamma$), the coordinates of $C_1$ and $C_2$ are given in Equations (\ref{RotateFinalOC_{1}}) and (\ref{RotateFinalOC_{2}}) in Appendix \ref{sec:appA}, to which interested readers may refer. Now, we are able to discuss the conditions required for a spacecraft traversing the CS.

\subsection{Three Criteria for Spacecraft Traversing a Given CS} \label{sec:Criteria}

Obviously, the premise for a spacecraft to traverse the CS is that its orbit should intersect the CS. Alternatively, to tell whether a CS could be traversed by a spacecraft needs to answer these two questions: (1) What kind of CS will intersect a given orbit, or conversely, what kind of orbit will intersect a CS whose position and morphology are known? (2) If the intersection of the orbit and the CS is realized, how could the spacecraft traverse the CS?

\begin{figure}[ht]
\begin{center}
\includegraphics[width=0.7\textwidth]{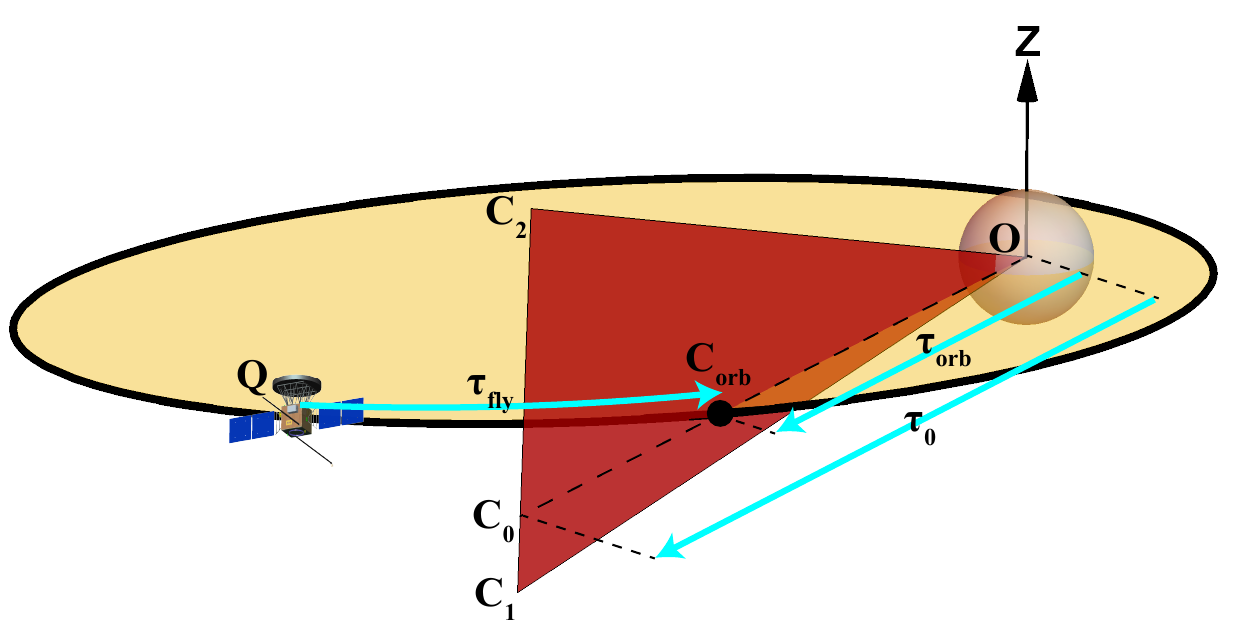}
\end{center}
\caption{The figure depicts a conceptual representation of a CS that may be traversed by a spacecraft. The orbital plane intersects the CS at the line $OC_0$, while the orbit itself crosses the CS at the point $C_{orb}$. $C_1$ and $C_2$ mark two endpoints of the CS. $Q$ is the location of the spacecraft at the eruption time $t$. The three cyan arrows represent the three characteristic periods: $\tau_0$, $\tau_{orb}$ and $\tau_{fly}$, which indicate the lifetime of the CS, the time of the CS propagation to the orbit, and the time of the spacecraft flight from $Q$ to $C_{orb}$, respectively.
\label{fig:Criterion}}
\end{figure}

First of all, we need to establish the criterion for determining the intersection of the orbit and the CS. Figure \ref{fig:Criterion} displays how the intersection occurs. The yellow ellipse is the area surrounded by the orbit and the red triangle is for the CS. For an infinitely long CS, namely $l\gg R_{\odot}$, the intersection occurs as $C_1(x_1,y_1,z_1)$ and $C_2(x_2,y_2,z_2)$ are located on either side of the orbital plane so that:
\begin{eqnarray}
Criterion~1:~z_1z_2<0\label{Criterion1}
\end{eqnarray}
with $z=0$ being the orbital plane that intersects the CS at line $OC_{orb}$, which extends to point $C_{0}$ located at line $C_{1}C_{2}$. After simple algebraic calculations, we have
\begin{eqnarray}
{\bf{OC_0}}=\left[x_0,y_0,z_0\right]^\mathrm{T}=\left[\frac{x_2z_1-x_1z_2} {z_1-z_2},\frac{y_2z_1-y_1z_2}{z_1-z_2},0\right]^\mathrm{T}\label{C_0}.
\end{eqnarray}
Criterion 1 suggests that in order for a spacecraft to traverse the CS, we need to pay more attention to the case in which the eruption takes place near the orbital plane, the resultant CME propagates in the direction roughly parallel to the orbital plane, and the associated CS is nearly orthogonal to the orbital plane as shown in Figure \ref{fig:Criterion}. Otherwise, the probability of the intersection is fairly low.

Second, for a CS with a finite length, say $l < 100R_{\odot}$, Criterion 1 is not strong enough to finalize the condition for intersection, and the impact of the finite value of $l$ on the criterion of intersection needs considering. Obviously, as Criterion 1 is satisfied, the CS intersects the orbit only if point $C_0$ lies outside the elliptical orbit, which yields to the second criterion:
\begin{eqnarray}
Criterion~2:~\frac{(x_0-c)^2}{a^2}+\frac{y_0^2}{b^2}>1.\label{Criterion2}
\end{eqnarray}
It suggests that the CS needs to develop apparently in length before it can intersect the orbit.

These two criteria clearly illustrate the conditions required for the intersection of the orbit and CS. We now discuss the condition necessary for a spacecraft to traverse the CS. In Figure \ref{fig:Criterion}, let $t$ be the eruption time, and $Q$ be the spacecraft position at time $t$. The cyan arrows represent three critical time intervals: $\tau_{orb}$, $\tau_{0}$, and $\tau_{fly}$, which demonstrate the time needed for the CS to propagate to point $C_{orb}$, at which the orbit intersects the CS, from the solar surface, the lifetime of CS (see the discussions that will be given shortly), and the time that the spacecraft takes to fly from point $Q$ to point $C_{orb}$, respectively.

On one hand, the CS is continually dissipated by magnetic reconnection, so the spacecraft must be located at a position not very far from the CS in order to cross the CS before the CS disappears. On the other hand, the spacecraft should not be very close to the plane where the CS is supposed to be, otherwise, the spacecraft may pass $C_{orb}$ before the developing CS touches the orbit and miss the chance to traverse the CS. Combining these two considerations yields the third criterion required for the spacecraft to traverse the CS:
\begin{eqnarray}
Criterion~3:~\tau_0>\tau_{fly}>\tau_{orb}\label{Criterion3}.
\end{eqnarray}
The first time interval, $\tau_0$, could be obtained according to observations \citep[][]{2016SoPh..291.3725W}, and we set
\begin{eqnarray}
\tau_0=18~\text{hrs}\label{tau_0}
\end{eqnarray}
throughout this work (see more discussions in Section \ref{sec:Finite case}). The second time interval, $\tau_{fly}$, is given as:
\begin{eqnarray}
\tau_{fly}=t_{orb}-t\label{tau_fly},
\end{eqnarray}
where $t_{orb}$ and $t$ are the times for the spacecraft to travel from the perihelion to points $C_{orb}$ and $Q$, respectively.
Interested readers refer to Appendix \ref{sec:appB} for more details. Here, we assume that the CS develops at a constant speed, $v$, then the third time interval, $\tau_{orb}$, is given by:
\begin{eqnarray}
\tau_{orb}=\frac{OC_{orb}} {v}\label{tau_orb},
\end{eqnarray}
where $OC_{orb}=a(1-e^2)/(1-e\cos{\phi})$, $e=c/a$, and $\phi=\arctan{(y_0/x_0)}$. \cite{2019SSRv..215...39L} investigated the correlation of the velocity to the acceleration of CME. They found the correlation is poor and the average acceleration is almost $0$, which suggests that the CS increases in length at a constant speed is not a bad approximation.

We realize that Criteria 1 and 2 impose constrictions on the size and orientation of the orbit and the CS, as well as the location of the source region of the eruption; and that Criterion 3 requires that the spacecraft motion and the CS kinematic behaviors to be further constricted. In this work, we name the CSs that satisfy Criteria 1 and 2 the ``candidates of detectable CSs (CDCSs)''. Obviously, satisfying Criterion 3 allows a CDCS to be a ``detectable CS (DCS)''. We are now ready to apply these 3 criteria to determine whether a given CS is detectable.

As mentioned earlier, we assume the half angular width of the CS $\delta=23^\circ$ in this work. The lifetime, $\tau_0$, is
given as $18$~hrs (see Equation \ref{tau_0} and relevant discussions in Section \ref{sec:Finite case}), and the CS length, $l$, is related to its lifetime, $\tau_{0}$, and extending velocity, $v$, in the way $l=v\tau_{0}$. Therefore, the CS is characterized by five parameters: $\theta_s$, $\phi_s$, $\gamma$, $l$, and $t$. According to the discussions above, the parameters that govern Criterion 1 include the orbital parameters, the rotational axis $rot$, and the inclination $\alpha$, as well as the CS parameters, $\theta_s$, $\phi_s$, and $\gamma$. The parameters that govern Criterion 2 include the orbital parameters, $a$, $b$, and $rot$, as well as the CS parameters, $\theta_s$, $\phi_s$, $\gamma$, and $l$. Note that Criterion 1 only applies to infinitely long CS, thus parameters related to length do not affect the result, but the situation for Criterion 2 changes, and it is impacted by $a$, $b$, and $l$. In addition to all the parameters governing Criterion 2, Criterion 3 takes $t$ into account, which is related to the development of the CS and the motion of the spacecraft.

\subsection{Probability Model} \label{sec:Probability}

This section provides an introduction to estimating the probability of traversing CSs by a spacecraft. In the case where the spacecraft orbit is given, the orbital parameters, $a$, $b$, $c$, and $\alpha$ are fixed, and the results given by Equations (\ref{Criterion1}) through (\ref{Criterion3}) are governed by the parameters related to CS only. In the framework of the probability, an event $\left\{\theta_s,\phi_s,\gamma,l,t\right\}$ is said to occur when a CS with parameters $\theta_s, \phi_s, \gamma, l$ and $t$ is produced by a solar eruption that is considered happening randomly. Thus, the occurrence of this event is equivalent to obtaining the coordinates of a random point in a five-dimensional parameter space spanned over $(\theta_{s}, \phi_{s}, \gamma, l, t)$. Therefore, the event ``the parameters of CS meeting three criteria'' can be considered equivalent to another event, ``a random point located in a sub-domain of the parameter space''. In other words, the probability of spacecraft traversing the CS could be obtained via evaluating the probability of which a given point is found in a sub-domain of the space $(\theta_{s}, \phi_{s},  \gamma, l, t)$. Obviously, this sub-domain is subject to the restriction of the three criteria strictly.

As mentioned earlier, Criterion 1 defines a large domain in the space spanned by $(\theta_s, \phi_s, \gamma)$, and we denote this large domain as $\Omega_{1}$. Referring to Figure \ref{fig:Criterion}, we realize that the points located in $\Omega_{1}$ help select the CS that possesses both the right location and the right orientation that might probably allow the traversing to occur. Similarly, Criterion 2 defines a smaller sub-domain $\Omega_2$ in a four-dimensional parameter space spanned by $(\theta_{s}, \phi_{s}, \gamma, l)$, and helps select the CDCS. Finally, Criterion 3 defines the smallest sub-domain $\Omega_3$ in the five-dimensional parameter space spanned by $(\theta_{s}, \phi_{s}, \gamma, l, t)$, and determines whether the traversing could eventually occur.

For an infinitely long CS, if the parameters for its location and orientation, $(\theta_{s}, \phi_{s}, \gamma)$, are located in $\Omega_{1}$, namely $(\theta_{s}, \phi_{s}, \gamma) \in \Omega_{1}$, it is a CDCS, independent of $l$ and $t$. Therefore, the corresponding probability of an infinitely long CS being a CDCS, $P^{CD}_{\infty}$ is:
\begin{eqnarray}
P^{CD}_{\infty}=P\left[(\theta_s,\phi_s,\gamma) \in \Omega_1\right]\label{P define1}.
\end{eqnarray}
Here $P^{CD}_{\infty}$ is calculated by integrating the joint probability density function (PDF) $f(\theta_s,\phi_s,\gamma)$ over the domain $\Omega_1$:
\begin{eqnarray}
P^{CD}_{\infty}=\int_{\Omega_1}f(\theta_s,\phi_s,\gamma) {\rm d}\theta_s {\rm d}\phi_s {\rm d}\gamma \label{P Omega1}.
\end{eqnarray}

Similarly, for a finitely long CS described by $(\theta_s, \phi_s, \gamma, l) \in \Omega_{1,2}$, where $\Omega_{1,2}=\Omega_{1} \cap \Omega_2$, parameters of the CS satisfy Criterion 1 and 2 simultaneously, the CS is a CDCS with the corresponding probability, $P^{CD}$, written as:
\begin{eqnarray}
P^{CD}=P\left[(\theta_s, \phi_s, \gamma, l) \in \Omega_{1,2}\right]\label{P define2},
\end{eqnarray}
and can be evaluated via integrating the joint PDF, $f(\theta_s,\phi_s,\gamma,l)$ over domain $\Omega_{1, 2}$
\begin{eqnarray}
P^{CD}=\int_{\Omega_{1,2}}f(\theta_s,\phi_s,\gamma,l) {\rm d}\theta_s {\rm d}\phi_s {\rm d}\gamma{\rm d}l\label{P Omega2}.
\end{eqnarray}

For the same reason, over the domain $\Omega_ {1,2,3}=\Omega_{1} \cap \Omega_2 \cap \Omega_3$, the probability of a CS being traversed by a spacecraft, $P^{tra}$, is given as:
\begin{eqnarray}
P^{tra}=P\left[(\theta_s, \phi_s, \gamma, l, t) \in \Omega_{1,2,3}\right]\label{P define3},
\end{eqnarray}
and
\begin{eqnarray}
P^{tra}=\int_{\Omega_{1,2,3}}f(\theta_s,\phi_s,\gamma,l,t) {\rm d}\theta_s {\rm d}\phi_s {\rm d}\gamma{\rm d}l{\rm d}t\label{P Omega3}.
\end{eqnarray}

Now, the initial problem of evaluating the probability of a spacecraft traversing a CS has been transformed into completing integrals in Equations (\ref{P Omega1}), (\ref{P Omega2}) and (\ref{P Omega3}) over the domains determined by the three criteria (see Equations \ref{Criterion1}, \ref{Criterion2}, and \ref{Criterion3}). We employed the Monte Carlo method to numerically evaluate these integrals. For simplicity, we conducted the computation by sampling the points uniformly in the integration domain. We discuss how to close these integrals in the next section.

\section{Probability for the Orbits of Probes to Traverse a CS} \label{sec:Result interact}

In this section, we conduct a detailed analysis of the probability of various orbits intersecting CME-flare CSs. Investigations are performed for the cases of infinitely and finitely long CSs separately.

\subsection{Infinitely Long CS} \label{sec:Infinite case}

In reality, a CS could never be infinitely long; for the mathematical construction, on the other hand, we perform an investigation of the probability of the spacecraft traversing an infinitely long CS. For an infinitely long CS, as mentioned before, it is a CDCS as long as its parameters $(\theta_{s}, \phi_{s}, \gamma) \in \Omega_{1}$. Several issues related to $\Omega_{1}$ are worth paying extra attention. First, how many sample points are included in $\Omega_{1}$, which reveals the attribute of the CS that could be traversed by the spacecraft? Second, how large is the range of variations of parameters covered by these samples, which determines whether a CS is a CDCS? Third, how does a given orbit affect $\Omega_{1}$, which reveals the impact of orbit parameters on $\Omega_{1}$.

\begin{figure}[ht]
\begin{center}
\includegraphics[width=1.0\textwidth]{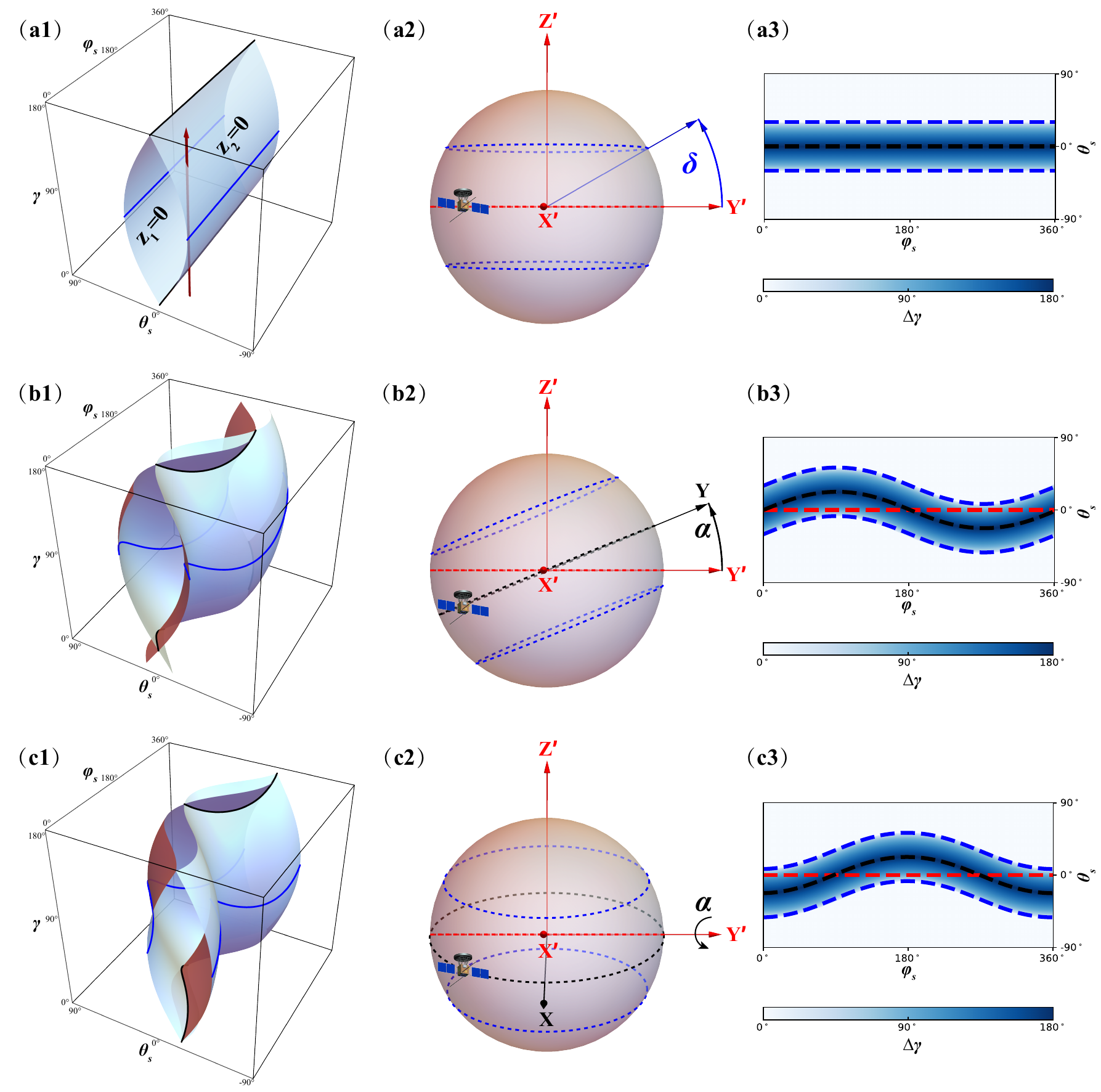}
\end{center}
\caption{The conditions required for an infinitely long CS to intersect with different orbits. From the top to the bottom panels, three rows correspond to the orbits that: (1) $\alpha=0^\circ$, (2) $\alpha \neq 0$ and $rot=X'$, and (3) $\alpha \neq 0$ and $rot=Y'$. Left panels depict the sub-domain $\Omega_1$ in the three-dimensional parameter space with $\theta_s$, $\phi_s$, and $\gamma$ as coordinates. The red arrow in panel (a1) represents the set of CSs that erupt at the source $(\theta_s, \phi_s)$, with $\gamma$ being allowed to take on any value. The blue lines in panels (a1-c1) represent the sets of tangent points of the red arrow to $\Omega_1$. In the medium panels, we show the locations that could generate CDCSs, which is the domain bounded by the two blue dashed lines. The red and black dashed lines indicate the equator and the projection of the orbit, respectively. The right panels display the $\Delta \gamma$ of CDCSs that erupt from different sources. The lines of different colors in the right panels have the same meaning as the medium panels.
\label{fig:Parameters_Space}}
\end{figure}

The left panels (a1), (b1), and (c1) in Figure \ref{fig:Parameters_Space} depict the sub-domain $\Omega_1$ in the space $(\theta_{s}, \phi_{s}, \gamma)$. They illustrate three distinct orbits: (1) $\alpha=0$, the orbital plane coinciding with the ecliptic plane (Figure \ref{fig:Parameters_Space}a1), (2) $\alpha \neq 0$, the orbital plane tilted away from the ecliptic plane by rotating an angle of $\alpha$ around the $X^{\prime}$-axis (Figure \ref{fig:Parameters_Space}b1), and (3) $\alpha \neq 0$, the orbital plane deviating away from the ecliptic plane by rotating the same angle around the $Y^{\prime}$-axis (Figure \ref{fig:Parameters_Space}c1). Since $\alpha \approx 0$ for PSP, Figure \ref{fig:Parameters_Space}a1 exhibits the case that is roughly suitable for PSP, in which $\Omega_1$ is bounded by two surfaces defined by $z_1=0$ and $z_2=0$ [see Equation (\ref{Criterion1}) and the related discussions]. The region of $z_{1}<z_{2}<0$ is located to the right of $\Omega_1$, the region of $z_{2}>z_{1}>0$ is to the left, and that of $z_{1}<0<z_{2}$ lies within $\Omega_1$.

The red arrow in Figure \ref{fig:Parameters_Space}a1, normal to the $\theta_{s}\phi_{s}$-plane with its tip located at surface $\gamma=0$, represents the CSs produced by an eruption from a fixed source $(\theta_s,\phi_s)$, while its orientation varies freely in a given range.

Figure \ref{fig:Parameters_Space}a1 indicates that if $(\theta_{s}, \phi_{s}, \gamma) \in \Omega_{1}$, the associated arrow touches $\Omega_{1}$, and the corresponding eruption would produce a CDCS. Figure \ref{fig:Parameters_Space}a1 indicates that the arrow and surface $z_{1}=0$ or $z_{2}=0$ could have two, one, and no intersection. In the case of two intersections, the eruption produces the CDCS as the value of $\gamma$ is in the range between these two intersections. We denote this range as $\Delta \gamma$.

As the arrow is tangential to either of the two surfaces, the locations of the arrow in the $\theta_{s}\phi_{s}$-plane set up the upper and lower boundaries of $\theta_{s}$ and $\phi_{s}$ (see two blue lines in Figure \ref{fig:Parameters_Space}a1), and no CDCS could be created outside these boundaries. These two boundaries are determined by the equations:
\begin{eqnarray}
z_i&=&0,~i=1,2,\label{Param_surface}\\
\frac{\partial \theta_s}{\partial \gamma}&=&0,\label{Param_curve}
\end{eqnarray}
from which we could eliminate $\gamma$ and obtain the equation of the two boundaries describing by $\theta_s$ and $\phi_s$:
\begin{eqnarray}
\cos{\alpha}\sin{\theta_s}-\sin{\alpha}\cos{\theta_s}\sin{\phi_s}&\equiv&\sin(\pm\delta),\label{Param_blueline}
\end{eqnarray}
where $\sin(+\delta)$ and $\sin(-\delta)$ correspond to $i=1$ and $i=2$ in Equation (\ref{Param_surface}), respectively.

In the case either that the arrow does not intersect surfaces $z_{1}=0$ and $z_{2}=0$, or that the value of $\gamma$ is outside the above range, no CDCS could be created as well.

In Figure \ref{fig:Parameters_Space}a2, the upper and lower boundaries of $\theta_{s}$ at the solar surface are outlined by two blue dashed curves. Only eruptions from regions on the solar surface between the two boundaries could produce CDCS. As we mentioned earlier, the $X^{\prime}Y^{\prime}$-plane is the ecliptic plane that is co-located in space with the orbital plane, $XY$-plane, in the case of $\alpha=0$. According to Equation (\ref{Param_blueline}), we obtain
\begin{eqnarray}
\theta_s\equiv\pm \delta\label{thetas_delta}
\end{eqnarray}
as $\alpha=0$ so the latitude of the blue dashed curves in Figure \ref{fig:Parameters_Space}a2 is identified with the half angular width of the CS, $\delta$. As expected, the larger the value of $\delta$ is, the larger of the source region on the solar surface that could produce the CME associated with the CDCS is.

Furthermore, the region between the two dashed curves in Figure \ref{fig:Parameters_Space}a2 corresponds to the straight blue belt in $\theta_{s}\phi_{s}$-plane in Figure \ref{fig:Parameters_Space}a3. The size of the belt gives the range of $\theta_{s}$ and $\phi_{s}$ for CDCS, and the color shading describes that of $\gamma$, namely $\Delta\gamma$. At the center of this belt, $\Delta\gamma$ attains its maximum, 180$^{\circ}$, indicating that the CS with any value of $\gamma$ falls under the category of CDCS. At the boundary of this belt, on the other hand, $\Delta\gamma$ vanishes, implying that any CS developing from this location and beyond is not detectable.

When $\alpha\neq 0$, the structure of $\Omega_1$ defined by Criterion 1 becomes complex, as depicted in Figures \ref{fig:Parameters_Space}b and \ref{fig:Parameters_Space}c. Specifically, Figures \ref{fig:Parameters_Space}b1 and \ref{fig:Parameters_Space}c1 illustrate the cases of which the orbital plane deviates from the ecliptic plane by rotating around the $X^{\prime}$- and the $Y^{\prime}$-axis, respectively. We repeat the analyses for the case of $\alpha=0$, apply the approach to the cases of $\alpha\neq 0$, and obtain the new detectable domains as shown in Figures \ref{fig:Parameters_Space}b2 and \ref{fig:Parameters_Space}c2.

To quantitatively describe the new domain in which the CS is detectable, we define $\theta$ as the latitude in the $XYZ$-system, which is related to $\theta_s$ and $\phi_s$ according to Equations (\ref{RotateFinalOC_{1}}) and (\ref{RotateFinalOC_{2}}):
\begin{eqnarray}
\cos{\alpha}\sin{\theta_s}-\sin{\alpha}\cos{\theta_s}\sin{\phi_s}&=&\sin{\theta}.\label{Param_theta}
\end{eqnarray}
Compared Equations (\ref{Param_blueline}) through (\ref{Param_theta}), we find
\begin{eqnarray}
\theta\equiv\pm \delta,\label{theta_delta}
\end{eqnarray}
which means that the range or scale of the domain in which the CS is detectable depends solely on $\delta$, and only the eruption from the region between latitude of $\theta=\delta$ and $\theta=-\delta$ can develop CDCS. The results suggest that the new detectable domain can be obtained by rotating the domain with the same range in Figure \ref{fig:Parameters_Space}a2 by an angle $\alpha$ around the $X'$- or the $Y'$-axis, coinciding with the way the orbital plane is rotated.

Figures \ref{fig:Parameters_Space}b3 and \ref{fig:Parameters_Space}c3 present the same information as Figure \ref{fig:Parameters_Space}a3 but for different cases of $\alpha\neq 0$. Comparing with the detectable domain that appears as a straight belt for the case of $\alpha=0$ (see Figure \ref{fig:Parameters_Space}a3), the regions in the $\theta_{s}\phi_{s}$-plane for $\Omega_{1}$ displayed in Figures \ref{fig:Parameters_Space}b3 and \ref{fig:Parameters_Space}c3 fluctuate periodically, and the greater the value of $\alpha$ is, the stronger the fluctuation is. Furthermore, a phase difference of $\pi/2$ exists between the cases in which the orbital plane deviates from the elliptic plane in different fashions.

Figures \ref{fig:Parameters_Space}b3 and \ref{fig:Parameters_Space}c3 reveal that the probability of producing a CDCS by the eruption from the same source varies with the way that the ecliptic plane deviates from the orbital plane. Specifically, for $rot=X'$, CMEs from the north hemisphere are more likely to produce CDCS when erupting from longitudes of $\phi_s \in (0^\circ, 180^\circ)$, whereas in the case of $rot=Y'$, the corresponding range of longitudes moves to $\phi_s \in (90^\circ, 270^\circ)$. As the CME occurs in the south hemisphere, on the other hand, the corresponding values of $\phi_{s}$ are just outside the above ranges. We also notice that positions with $\Delta\gamma = 180^{\circ}$ are mainly located around the intersection of the orbital plane and the solar surface (black dashed curve), instead of around the solar equator (red dashed curve). This implies that eruptions around the orbital plane are more likely to produce CDCS. Generally speaking, it is important to note that for an infinitely long CS, whether it is a CDCS depends not only on the location of the eruption source region on the solar surface but also on the parameters of the orbit. Consequently, the probability distributions of CDCS versus CS parameters will be considered in a more realistic manner to accurately calculate the probability of intersection between the CS and different orbits.

To utilize Equation (\ref{P Omega1}) for evaluating the probability $P^{CD}_{\infty}$ of an infinitely long CS being a CDCS, we first need to obtain the joint PDF $f(\theta_s,\phi_s,\gamma)$. It is reasonable to assume that $\phi_s$ of a CDCS is independent of $\theta_{a}$ and $\gamma$, namely the CS source region latitude and CS tilt angle. Therefore, the joint PDF of the three parameters is:
\begin{eqnarray}
f(\theta_s,\phi_s,\gamma)=f(\theta_s,\gamma)f(\phi_s), \label{Density_tpg}
\end{eqnarray}
where $f(\phi_s)$ is the PDF of $\phi_s$ and $f(\theta_s,\gamma)$ is the joint PDF of $\theta_s$ and $\gamma$. As a plausible approximation, the PDF of $\phi_s$ can be assumed uniform. Therefore,
\begin{eqnarray}
f(\phi_s)&=&\frac{1}{360}, \phi_s \in(0, 360)\label{Density_phi}.
\end{eqnarray}
As for the joint PDF $f(\theta_s,\gamma)$, the optimal calculation method is to examine the proportion of CDCSs for different values of $\gamma$ and $\theta_s$ according to observations. However, it is difficult to directly infer the tilt angle $\gamma$ from the observed CSs because of the limit to observations.

\cite{2009SoPh..256..111T} estimated the tilt angle $\gamma$ based on observations of CMEs. However, for some CMEs with complex leading edges, it is difficult to determine the tilt angle. In order to obtain enough data and deduce the dependence of $\gamma$ on $\theta_s$, we use the tilt angle and latitude of filament channels on the solar disk instead of $\gamma$ and $\theta_s$ for CSs. \cite{2015ApJS..221...33H} used the automatic identification method to study various parameters of filament channels during solar cycles 22 to 24, including their tilt angle and latitude on the solar disk. In this work, we are using their results.

\begin{figure}[ht]
\begin{center}
\includegraphics[width=0.7\textwidth]{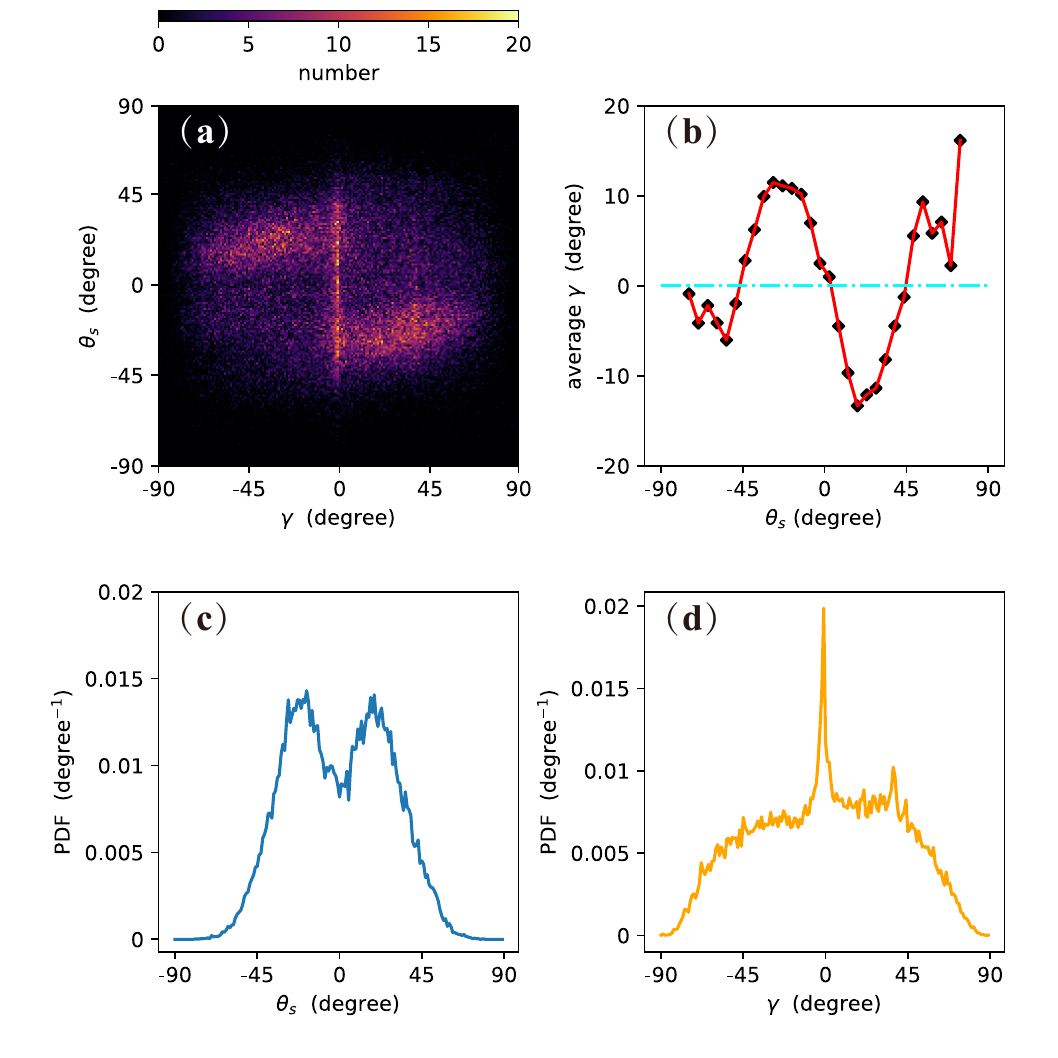}
\end{center}
\caption{Relationship between the tilt angle $\gamma$ and the latitude $\theta_s$. (a) The joint PDF of $f(\theta_s, \gamma)$ in the $(\theta_{s}, \gamma)$ space. The color represents numbers of the filament of different tilt angle, $\gamma$, at different latitudes, $\theta_{s}$ \citep[data from][]{2015ApJS..221...33H}. (b) Variations of the average $\gamma$ evaluated from panel (a) versus $\theta_s$. The marginal PDFs of $\theta_s$ and $\gamma$ are illustrated in panels (c) and (d).
\label{fig:pdf_theta_gamma}}
\end{figure}

Figure \ref{fig:pdf_theta_gamma}a shows the distribution of the filament numbers per unit area (in square degrees) at different latitudes and tilt angles. Dividing the values at each point in the figure by the total number of filaments yields the joint PDF $f(\theta_s,\gamma)$. We notice that a large number of filaments with $\gamma = 0$ distribute in a region over a big range of latitudes approximately from $-90^{\circ}$ to $90^{\circ}$. Figure \ref{fig:pdf_theta_gamma}b shows the value of average $\gamma$ at different latitudes. Information revealed from Figure \ref{fig:pdf_theta_gamma}b suggests that, on the north hemisphere, negative $\gamma$ dominates in the region near the equator, and the positive $\gamma$ dominates the high latitude region; on the other hand, the situation reverses on the south hemisphere, but the amplitude of the change in the mean $\gamma$ is higher in north than that in south. These results are consistent with the statistical results obtained by \cite{2016SoPh..291.1115T} on a longer time scale (about 100 years), confirming the reliability of the results obtained by \cite{2015ApJS..221...33H}. Furthermore, Figures \ref{fig:pdf_theta_gamma}a and \ref{fig:pdf_theta_gamma}b indicate that latitudes and orientations of the filament vary over a large range, but those orientations with relatively small $\lvert \gamma \rvert$ dominate.

According to Figure \ref{fig:pdf_theta_gamma}a, we are also able to obtain $f(\theta_{s})$ and $f(\gamma)$:
\begin{eqnarray}
f(\theta_{s}) &=& \int f(\theta_{s}, \gamma){\rm d}\gamma, \\
f(\gamma) &=& \int f(\theta_{s}, \gamma){\rm d}\theta_{s},
\end{eqnarray}
and Figures \ref{fig:pdf_theta_gamma}c and \ref{fig:pdf_theta_gamma}d display $f(\theta_s)$ and $f(\gamma)$, respectively. Figure \ref{fig:pdf_theta_gamma}c shows that the latitude distribution of filaments has a bimodal structure, with the most filaments appearing in the regions around latitudes of $\pm 30^{\circ}$. Figure \ref{fig:pdf_theta_gamma}d once again illustrates that the probability of filaments with $\gamma=0$ is the highest.

Overall, we obtained the joint PDF $f(\theta_s,\gamma)$ of filaments through observational data, and analyzed the correlation of $\gamma$ to $\theta_s$. The key point is that the value of $\gamma$ of most CSs is close to 0. The meaning of this point is twofold. First, the CSs that can be observed are usually developed in the eruption occurring on either east or west edge of the Sun, and most of them were observed edge-on (e.g, see discussions of \citealt{2003ApJ...594.1068K,2010ApJ...722..625K}; \citealt{2005ApJ...622.1251L,2007ApJ...658L.123L,2009ApJ...693.1666L,2015SSRv..194..237L}; and \citealt{doi:https://doi.org/10.1002/9781119324522.ch15}). Second, decreasing the inclination angle $\alpha$ increases the frequency of the spacecraft crossing the CS since the eruption is more likely to occur in the middle and low latitude region; on the other hand, the spacecraft on the orbit of large inclination angle has more opportunities to cross the CS with large angle, even with right angle, which will help us attain accurate information about the CS thickness (e.g., see \citealt{2015SSRv..194..237L} and \citealt{doi:https://doi.org/10.1002/9781119324522.ch15} for more discussions on the importance of such information). To obtain an orbit with the highest probability of detecting the CS, we need to balance the above two aspects regarding the inclination angle.

\begin{figure}[ht]
\begin{center}
\includegraphics[width=1.0\textwidth]{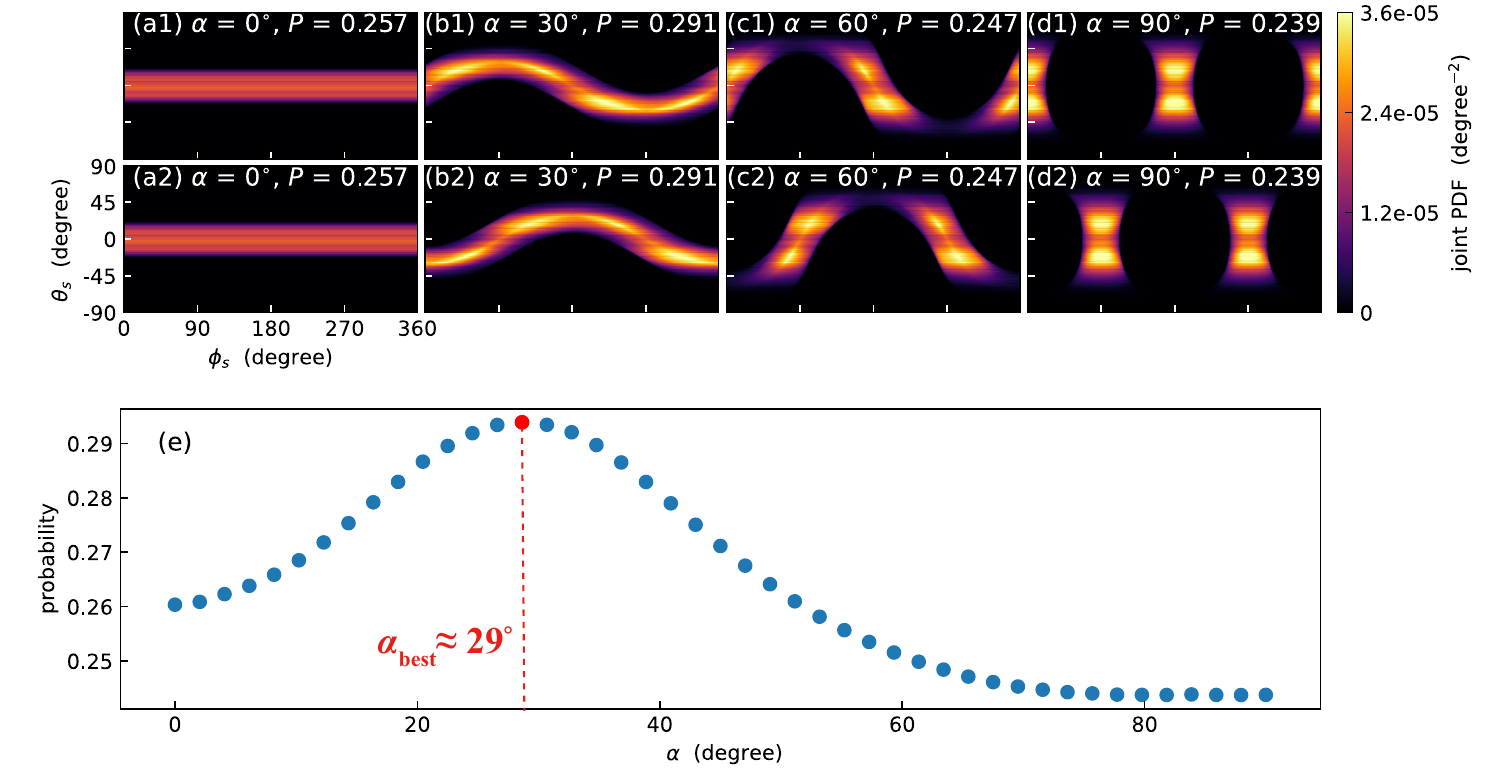}
\end{center}
\caption{Top two panels show the probability of different orbits intersecting an infinitely long CS erupted from a unit area in the solar surface, where $rot=X'$ in the first row and $rot=Y'$ in the second one. In the bottom panel, we display the intersection probability versus the inclination angle $\alpha$ of the orbit. The results indicate that the intersection probability is independent to $rot$ and the orbit with $\alpha=29^\circ$ is the most likely to intersect an infinitely long CS.
\label{fig:probability_alphas}}
\end{figure}

We employ a more realistic $f(\theta_s,\gamma)$ to evaluate the probability of a random CS being a CDCS for various orbits. We introduce the function $B(\theta_s,\phi_s)$, which describes the likelihood of generating a CDCS by an eruption from a unit area at location $(\theta_s, \phi_s)$ on the solar surface, and is given by:
\begin{eqnarray}
B(\theta_s,\phi_s)=\int\limits_{\Omega_1}f(\theta_s,\phi_s,\gamma){\rm d}\gamma \label{belt}.
\end{eqnarray}
The top two rows of Figure \ref{fig:probability_alphas} illustrate $B(\theta_{s}, \phi_{s})$ for different orbits (refer to Figure \ref{fig:Parameters_Space} for further clarification). This probability clearly depends on the location $(\theta_{s}, \phi_{s})$ where the eruption takes place. Figures \ref{fig:probability_alphas}a1 through \ref{fig:probability_alphas}d1 and \ref{fig:probability_alphas}a2 through \ref{fig:probability_alphas}d2 demonstrate the likelihood of an orbit intersecting the CS after deviating from the ecliptic plane at different angles of $\alpha$ by rotating the coordinate system around the $X'$- and $Y'$-axes, respectively. Similar to the results presented in Figures \ref{fig:Parameters_Space}c1 through \ref{fig:Parameters_Space}c3, parameter $rot$ determines the phase of the undulations of the belt region, while $\alpha$ controls their amplitudes. However, in contrast to Figure \ref{fig:Parameters_Space}, which only describes $\Delta \gamma$ at various latitudes and longitudes as a qualitative description of the intersection probability, Figure \ref{fig:probability_alphas} directly gives the probability. For the special cases, say $\alpha=90^\circ$, the CDCS could develop in the eruption from two regions. For $rot=X'$ (Figure \ref{fig:probability_alphas}d1), eruptions occurring near $\phi_s=0^\circ$ or $\phi_s=180^\circ$ are more likely to generate a CDCS, while for $rot=Y'$ (Figure \ref{fig:probability_alphas}d2), eruptions occurring near $\phi_s=90^\circ$ or $\phi_s=270^\circ$ are more favorable for producing CDCS. This further highlights that eruptions near the orbital plane are more likely to generate CDCSs. In addition, we can also calculate $P^{CD}_{\infty}$ by integrating $B(\theta_s,\phi_s)$ to further investigate this phenomenon:
\begin{eqnarray}
P^{CD}_{\infty}=\int B(\theta_s,\phi_s){\rm d}\theta_s {\rm d}\phi_s \label{P_belt1}.
\end{eqnarray}
Figure \ref{fig:probability_alphas}e presents variations of the intersection probability of the orbit and an infinitely long CS versus the inclination angle $\alpha$. It is apparent that $\alpha$ affects this probability in the case of an infinitely long CS, whereas $rot$ does not. We find that as $\alpha$ increases, $P^{CD}_{\infty}$ initially increases and then decreases, with a peak value of $P^{CD}_{\infty}=0.29$ achieved at $\alpha \approx 29^\circ$. For the PSP orbit, the probability of an infinitely long CS being a CDCS is $P^{CD}_{\infty}=0.26$.

\subsection{Finitely Long CS} \label{sec:Finite case}

In reality, the length of CSs is finite. \cite{2016SoPh..291.3725W} studied about 130 CME-flare CSs in the solar maximum and the minimum of the 23rd solar cycle. They found that the average lengths of CSs in the maximum and the minimum years were $12.4R_{\odot}$ and $11.8R_{\odot}$, respectively. The longest CSs found so far were $18.5R_{\odot}$ and $17R_{\odot}$ long in the maximum and the minimum. Moreover, the average velocities of the CS increase in length during the solar maximum and the minimum years were $324$~km~s$^{-1}$ and $188$~km~s$^{-1}$, respectively, with the corresponding accelerations of $6.3$~m~s$^{-2}$ and $8.3$~m~s$^{-2}$, respectively. The average lifetimes of CSs during the maximum and the minimum years were $\tau_0=16$~hrs and $\tau_0=18.2$~hrs, respectively. Assuming that the CS extends with a constant velocity, then according to the mean velocity and lifetime of CSs from observations, we are able to deduce the mean length of the CS in the maximum and the minimum years to be approximately $27R_{\odot}$ and $18R_{\odot}$, respectively. We note here that, to our knowledge for the time being, no report has ever been given about the true length of the CME-flare CS so far, and the longest CS that so far was reported and could be definitely identified is the one observed by LASCO/C3, which is between 20~R$_{\odot}$ and 30~R$_{\odot}$ \citep[e.g., see][]{2005ApJ...622.1251L}. We understand that, due to the limit of the observational techniques to our capabilities of acquiring the complete physical scenario regarding the CS, the length of a CME-flare CS should be longer than what we have known. This is also true for the lifetime of the CS. Therefore, both the length and the lifetime of the CME-flare CS used in the present work might be just a lower limit to the true values of the CS in reality. Hence, 27~R$_{\odot}$ and 18~R$_{\odot}$ for the CS length are used as references in the present work. The relevant parameters mentioned above are summarized in Table \ref{tab:CSs_kinematic}.
\begin{deluxetable*}{c|cccccc}[h]
\tablenum{1}
\tablecaption{Parameters of CSs according to \cite{2016SoPh..291.3725W}.\label{tab:CSs_kinematic}}
\tablewidth{0pt}
\tablehead{
\colhead{} &\colhead{Maximum Year} & \colhead{Minimum Year}
}
\startdata
Average $l$~($R_{\odot}$) & 12.4 & 11.8\\
Longest $l$~($R_{\odot}$) & 18.5 & 17\\
Average $v$~(km~s$^{-1}$) & 324 & 188\\
Average acceleration~(m~s$^{-2}$) & 6.3 & 8.3\\
Average $\tau_0$~(hrs) & 16 & 18.2\\
Estimated $l = v\tau_{0}$~($R_{\odot}$) & $\approx 27$ & $\approx 18$\\
\enddata
\end{deluxetable*}
\cite{2016SoPh..291.3725W} pointed out that since the CS is gradually dissipated, the estimated lifetime $\tau_0$ is just a lower limit, because the fact that the CS disappears from observational data does not necessarily mean that it does not exist any longer, but only means that its emission measure in the given wavelength is below the sensitivity of the detector. \cite{2013ApJ...766...65C} even identified a CS with a lifetime of approximately 38~hrs when analyzing the white-light data from LASCO. Recent observations of the Wide-Field Imager for Solar Probe \citep[WISPR;][]{2016SSRv..204...83V} onboard the PSP showed that more complex CS structures were seen in the white light and their durations are longer than those observed near the Earth when the probe is very close to the Sun \citep[][]{2022ApJ...936...43H}.

Using Equation (\ref{P Omega2}), we can calculate the probability $P^{CD}$ that a finitely long CS is a CDCS. For the joint PDF $f(\theta_s, \phi_s, \gamma, l)$, as mentioned before, the dependence of $f(\theta_{s}, \phi_{s}, \gamma, l)$ on $\phi_{s}$ does not correlate to the dependence of $f$ on any other variables such that
\begin{eqnarray}
f(\theta_{s}, \phi_{s}, \gamma, l) = f(\theta_{s}, \gamma, l) f(\phi_{s}).
\end{eqnarray}
However, obtaining the joint PDF of the variables $(\theta_s, \gamma, l)$ is still difficult due to the lack of sufficient statistical samples on latitude, inclination, and length of CSs. Therefore, we make a relatively strong assumption that the length of the CS, $l$, is also independent of the other variables. Thus, we obtain:
\begin{eqnarray}
f(\theta_s,\phi_s,\gamma,l)=f(\theta_s,\gamma)f(\phi_s)f(l) \label{Density_tpgl},
\end{eqnarray}
where $f(l)$ is the marginal density of $l$ and describes the likelihood that a CS with length of $l$ occurs. Combining Equations (\ref{P Omega2}) and (\ref{Density_tpgl}), $P^{CD}$ can be expressed as:
\begin{eqnarray}
P^{CD}&=&\int_{\Omega_{1,2}}f(l)f(\theta_s,\gamma)f(\phi_s) {\rm d}\theta_s {\rm d}\phi_s {\rm d}\gamma{\rm d}l,\nonumber \\
&=&\int f(l) \left[\int_{\Omega_{1,2}^l}f(\theta_s,\gamma)f(\phi_s){\rm d}\theta_s {\rm d}\phi_s {\rm d}\gamma \right]{\rm d}l,\label{P Omega2.2}
\end{eqnarray}
where $\Omega_{1,2}^l$ is the sub-domain inside $\Omega_{1,2}$ for a given $l$. We define $P^{CD}_l$ as the conditional probability, which quantifies the likelihood of a CS being CDCS as $l$ is known. Consequently, according to the law of the total probability, we obtain:
\begin{eqnarray}
P^{CD}&=&\int f(l)P^{CD}_l{\rm d}l\label{P total},
\end{eqnarray}
combining Equations (\ref{P Omega2.2}) and (\ref{P total}) leads to:
\begin{eqnarray}
P^{CD}_l&=&\int_{\Omega_{1,2}^l}f(\theta_s,\gamma)f(\phi_s){\rm d}\theta_s {\rm d}\phi_s {\rm d}\gamma\label{P condition}.
\end{eqnarray}
\cite{2016SoPh..291.3725W} studied behaviors of the CME-flare CS comprehensively and revealed important information on $l$, but it is not enough to construct $f(l)$ because their samples include only 52 CSs. Instead, we look into the probability of a CS with the given length to be a CDCS, $P^{CD}_{l}$ in this part of work. We shall further investigate the influence of $f(l)$ on the final results later.

\begin{figure}[ht]
\begin{center}
\includegraphics[width=1.0\textwidth]{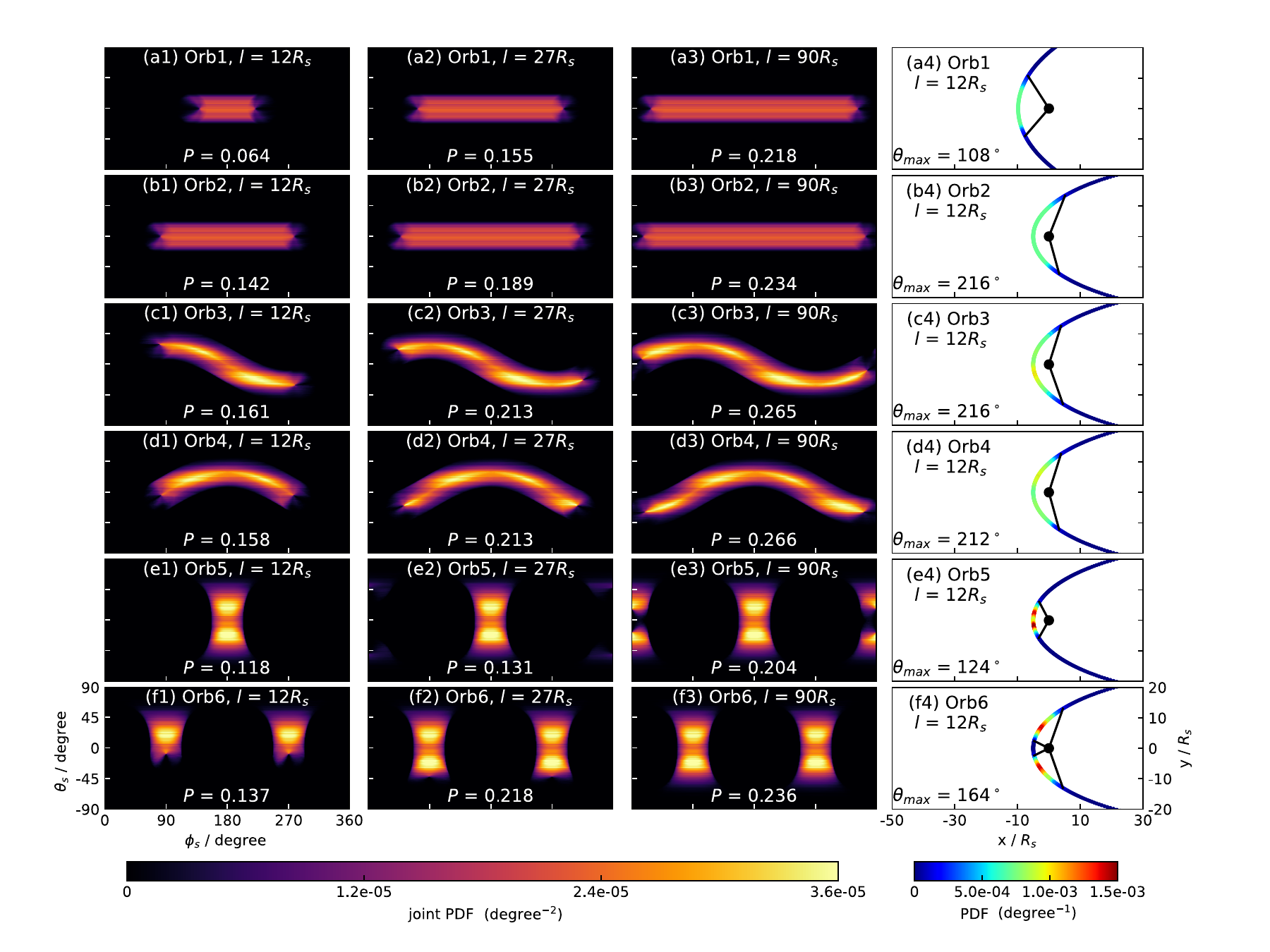}
\end{center}
\caption{The left three panels show the probability of various orbits intersecting a finitely long CS erupted from unit area in the solar surface, with the CS length of $l=12R_{\odot}, 27R_{\odot}$, and $90R_{\odot}$, respectively. The rightmost panel displays the intersection probability at different locations in the orbit. Panels through (a) to (f) correspond to the orbits from Orb$_1$ to Orb$_6$ listed in Table \ref{tab:cases_orb}.
\label{fig:probability_param_length}}
\end{figure}

We now demonstrate the probabilities of CSs being CDCSs when their lengths are $12R_{\odot}$, $27R_{\odot}$, and $90R_{\odot}$, respectively. These lengths are the average CS length obtained from LASCO data \citep[][]{2016SoPh..291.3725W}, the product of the average speed and the lifetime of CSs in the solar maximum \citep[][]{2016SoPh..291.3725W}, and the length of a hypothetically ultra-long CS that we may imagine. To perform further studies about the probability of the spacecraft crossing the CS, we consider six types of orbits (see Table \ref{tab:cases_orb}), among which Orb$_1$ is the PSP orbit, and Orb$_2$ is the orbit obtained by scaling down the PSP orbit. The left three columns in Figure \ref{fig:probability_param_length} show the detection probability belts, which represent the probability of producing CDCSs by the eruption from a unit area at the location $(\theta_{s}, \phi_{s})$ on the solar surface, $B_{l}(\theta_s,\phi_s)$. $B_{l}(\theta_s,\phi_s)$ is given as:
\begin{eqnarray}
B_{l}(\theta_s,\phi_s)=\int\limits_{\Omega_{1,2}^l}f(\theta_s,\gamma)f(\phi_s){\rm d}\gamma \label{belt2}.
\end{eqnarray}
Different from the probability given by Equation (\ref{belt}), the probability discussed here depends not only on $\theta_s$ and $\phi_s$, but on the length of the CS, $l$, as well.
\begin{deluxetable*}{c|cccccc}[h]
\tablenum{2}
\tablecaption{Parameters for different orbits.\label{tab:cases_orb}}
\tablewidth{0pt}
\tablehead{
\colhead{} &\colhead{Orb$_1$} & \colhead{Orb$_2$} & \colhead{Orb$_3$} & \colhead{Orb$_4$} & \colhead{Orb$_5$} & \colhead{Orb$_6$}
}
\startdata
$a$~$(R_{\odot})$ & 82.9 & 65 & 65 & 65 & 65 & 65\\
$b$~$(R_{\odot})$ & 39.1 & 25 & 25 & 25 & 25 & 25\\
$c$~$(R_{\odot})$ & 73.1 & 60 & 60 & 60 & 60 & 60\\
$perihelion$~$(R_{\odot})$ & 9.8 & 5 & 5 & 5 & 5 & 5\\
$aphelion$~$(R_{\odot})$ & 156 & 125 & 125 & 125 & 125 & 125\\
$rot$ & - & - & $X'$ & $Y'$ & $X'$ & $Y'$\\
$\alpha$~(degree) & 3.4 & 3.4 & 30 & 30 & 90 & 90\\
\enddata
\end{deluxetable*}

We first investigate the case where the orbital plane nearly coincides with the ecliptic plane ($\alpha \approx 0$, or the $XYZ$-system is identified with the $X^{\prime}Y^{\prime}Z^{\prime}$-system). Comparing Figure \ref{fig:probability_alphas}a1 with Figures \ref{fig:probability_param_length}a1 through \ref{fig:probability_param_length}a3 indicates that the size of the source region of the eruption that can produce CDCS significantly shrinks if the CS length is finite. Figure \ref{fig:probability_param_length} shows that, for a CS of $l=12R_{\odot}$, its intersection with the orbit is confined to a range of approximately $\Delta \theta=108^\circ$ near the perihelion. Comparing panels in Figures \ref{fig:probability_param_length}a1 through \ref{fig:probability_param_length}a3 with those in Figures \ref{fig:probability_param_length}b1 through \ref{fig:probability_param_length}b3 suggests that, for a CS that is not very long, intersection is more likely to occur with a small orbit. For example, when $l=12R_{\odot}$, the probability of Orb$_2$ crossing the CS ($P_{12R_{\odot}}^{CD}=0.142$) is two times that of the PSP orbit ($P_{12R_{\odot}}^{CD}=0.064$), and the corresponding range of $\theta$ in which Orb$_{2}$ could intersect with the CS is two times that of the PSP orbit intersects the CS ($216^{\circ} : 108^{\circ}$. See Figures \ref{fig:probability_param_length}a4 and \ref{fig:probability_param_length}b4). When $l=90~R{\odot}$ (see Figures \ref{fig:probability_param_length}a3 and \ref{fig:probability_param_length}b3), on the other hand, we find that $P_{90R_{\odot}}^ {det}=0.218$ for the PSP orbit, and $P_{90R_{\odot}}^ {det}=0.234$ for Orb$_2$. Therefore, for a CS that is not very long, the probability a small orbit crosses it is higher than that a large orbit crosses it. As the CS length increases, the difference in probabilities among different orbits crossing the CS decreases. As the CS becomes infinitely long, the difference vanishes.

Now we are looking into the case of $\alpha \neq 0$. For the orbits with $rot=X'$ (Orb$_3$ and Orb$_5$), the longer the CS is, the wider the longitude range is in which CDCSs occur. This is because the perihelion is located at $\phi_s=180^\circ$, as $l$ decreases, the region where CDCSs exist shrinks towards $\phi_s=180^\circ$, which is consistent with the cases of Orb$_1$ and Orb$_2$. For the orbits with $rot=Y'$ (Orb$_4$ and Orb$_6$), on the other hand, the latitude range of CDCS sources becomes wider as $l$ increases. This is because the perihelion of the orbit is located above the north hemisphere of the Sun, and it is difficult for the CS developing in the south to reach the orbit. Therefore, as $l$ decreases, the region where CDCS exists shrinks towards higher latitudes in the north hemisphere. These results indicate that, for the CS of finite length, only requiring the eruption to occur in the direction roughly parallel to the orbital plane could not help increase the probability of traversing. Apparently, it further requires the eruption to take place in the region near the perihelion.

We further analyze the case in which the orbital plane is orthogonal to the ecliptic plane (Orb$_5$ and Orb$6$). Our results, presented in Figures \ref{fig:probability_param_length}e1 and \ref{fig:probability_param_length}f1, reveal that comparing with the case of an infinitely long CS (Figures \ref{fig:probability_alphas}d1 and \ref{fig:probability_alphas}d2), the range of CDCS sources for Orb$_5$ and Orb$_6$ decreases by nearly half when $l=12~R{\odot}$ as a result of that Orb$_{5}$ lacks contribution in the longitudinal direction from the region near the aphelion (ascending node), while Orb$_{6}$ lacks contribution from the region extending in the latitudinal direction around the ascending and descending nodes above the south hemisphere. When $l=27~R{\odot}$ (Figures \ref{fig:probability_param_length}e2 and \ref{fig:probability_param_length}f2), the range of the CDCS parameters for Orb$_5$ is still concentrated around $\phi_s=180^\circ$, but Orb$_{6}$ could detect a large number of CSs produced by eruptions in the south hemisphere, and the probability of which Orb$_{6}$ intersects with CSs ($P_{27~R{\odot}}^{CD}=0.218$) is almost twice that of Orb$_5$ ($P_{27~R{\odot}}^{CD}=0.131$). This is because the ascending node of Orb$_5$ is very far from the Sun, making it difficult in detecting the CS even if $l$ increases. On the other hand, the ascending and the descending nodes of Orb$_6$ are not very far, and the intersecttion with the CS could occur around both locations. Figures \ref{fig:probability_param_length}e4 and \ref{fig:probability_param_length}f4 confirm this point quantitatively, and also indicate that opportunities for Orb$_5$ concentrates around the perihelion. Finally, for $l=90$~R${\odot}$, the probability of which the two orbits intersect the CS is roughly the same.

\begin{figure}[ht]
\begin{center}
\includegraphics[width=1.0\textwidth]{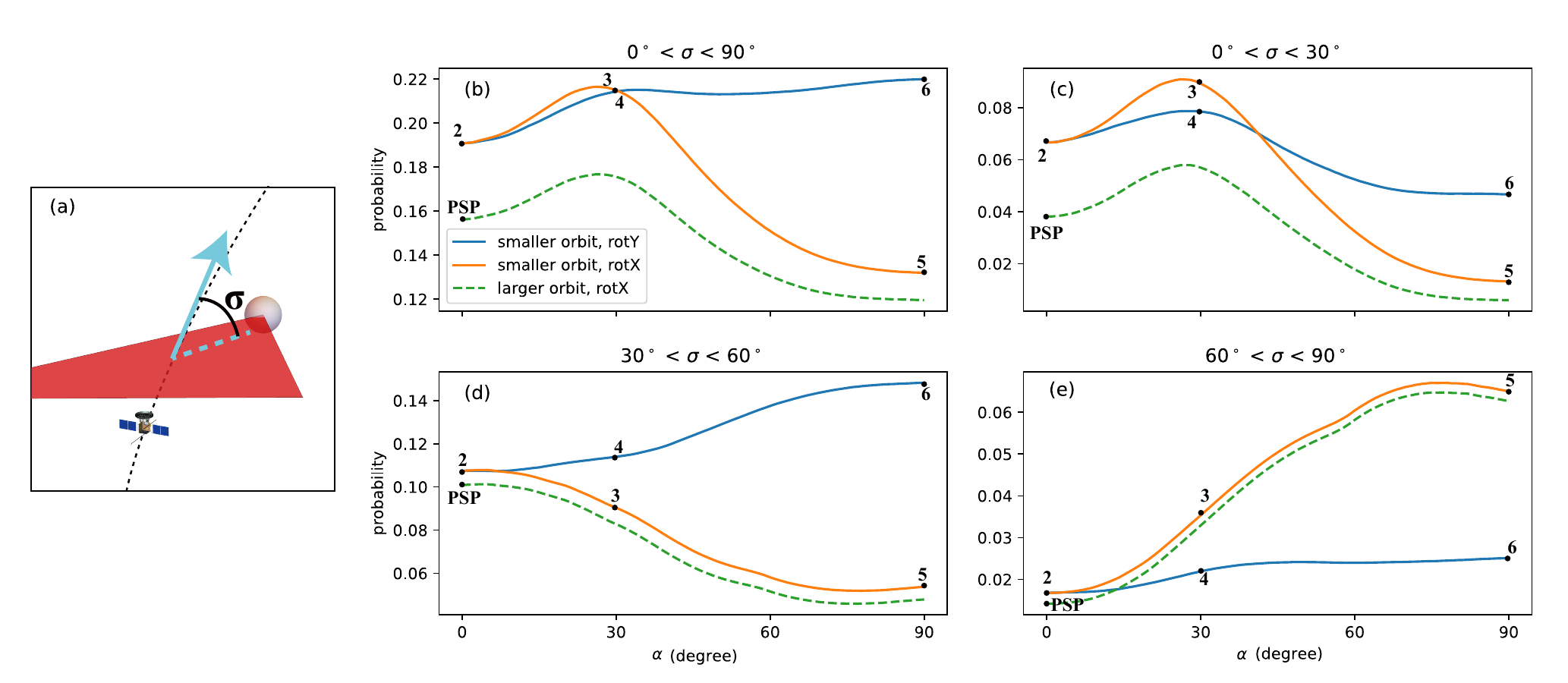}
\end{center}
\caption{(a) A sketch of the angle between the CS plane (red triangle) and the instantaneous velocity (cyan vector) of the spacecraft at the traverse moment, which is denoted as $\sigma$. Panels (b-e) display the intersection probabilities versus the inclination angle of orbit, and correspond to four situations that $\sigma$ could be any value, small $\sigma$, medium $\sigma$ and large $\sigma$. The blue and orange lines respectively correspond to the smaller orbits with $rot=Y'$ and $X'$, and the green dashed line matches the larger orbit with $rot=X'$, where the larger and smaller orbits means $a=82.9R_{\odot}, b=73.1R_{\odot}$ and $a=65R_{\odot}, b=60R_{\odot}$. The numbers in panels (b-e) correspond to different orbits in table \ref{tab:cases_orb}.
\label{fig:probability_sigma}}
\end{figure}

In addition, we further investigate the fashion in which they intersect, namely probability that they intersect at different angles, which is defined as $\sigma$. As shown in Figure \ref{fig:probability_sigma}a, $\sigma$ is the angle between the direction of the spacecraft motion and the plane of the CS at the intersection point. Since traversing the CS from either side has the same effect, we will not distinguish the ``front'' or the ``back'' side of CS, and only consider the case of which $0^\circ<\sigma<90^\circ$. We define the case of  $0^\circ<\sigma<30^\circ$ as the small angle traverse, that of $30^\circ<\sigma<60^\circ$ as the medium angle traverse, and that of $60^\circ<\sigma<90^\circ$ as the large angle traverse. Figures \ref{fig:probability_sigma}b  through \ref{fig:probability_sigma}e show the probability of the CS with $l=27R_{\odot}$ intersecting various orbits at different angles $\sigma$. Blue solid line, orange solid line, and green dashed line respectively represent three orbits: 1. $a=65R_ {\odot}$, $b=60R_{\odot}$, $rot=Y'$; 2. $a=65R_{\odot}$, $b=60R_{\odot}$, $rot=X'$; 3. $a=82.9R_{\odot}$, $b=73.1R_{\odot}$, $rot=X'$.

Figure \ref{fig:probability_sigma}b displays variations of probabilities versus $\alpha$ for all $\sigma$ between 0$^{\circ}$ and 90$^{\circ}$. We notice that, first of all, regardless of the value of $\alpha$, smaller orbits are more likely to encounter the CS than larger ones; second, for $rot=X'$, the value of $\alpha$ that leads to the highest probability of intersection is about $30^\circ$; third, for the orbit of $rot=Y'$, as $\alpha$ increases, the probability slightly increases and reaches its maximum at $\alpha=90^\circ$; fourth, for most of values of  $\alpha$, the orbit of $rot=Y'$ is more likely to encounter the CS than the orbit of $rot=X'$; and finally, eruptions near the ascending and the descending nodes of the orbit are more likely to produce CDCSs. However, the ascending node of the $rot=X'$ orbit is the aphelion, while the ascending and the descending nodes of the $rot=Y'$ orbit are much closer to the Sun, resulting in a parameter space for eruptions producing the CDCS almost twice that of the former (comparing Figures \ref{fig:probability_param_length}e2 and \ref{fig:probability_param_length}f2).

When considering the impact of individual $\sigma$, we notice that the probability of the spacecraft passing through the CS at a medium angle $\sigma$ is relatively high. As $\alpha$ increases, the probability profile of different orbits intersecting the CS exhibits different varying patterns. For the three orbits discussed above, probabilities of traversing the CS at small angles show an increasing-decreasing trend (see Figure \ref{fig:probability_sigma}c). The probability for the orbit of $rot=X'$ intersecting the CS at medium angles continues to decrease with $\alpha$, while that for the orbit of $rot=Y'$ crossing the CS slightly increases (see Figure \ref{fig:probability_sigma}d). The probability for the $rot=X'$ orbit intersecting the CS at large angles slightly increases, while that for the $rot=Y'$ situation increases at an almost negligible rate (see Figure \ref{fig:probability_sigma}e).

The above results indicate that the overall probability of the PSP orbit intersecting the CS is low, and it is difficult for the spacecraft to pass through the CS at large angles. The probabilities of PSP orbit crossing the CS at small and medium angles are 0.04 and 0.1, respectively, which seems fairly low but not impossible. The intersection probability of Orb$_2$ is higher than that of the PSP orbit, but it is mainly contributed by the case of small angle intersections. The case Orb$_3$ has the highest probability of encountering the CS at small angles among all the orbits. The intersection probability of Orb$_4$ is also high with contribution mainly from the medium angle intersections. The case Orb$_5$ belongs to the case of small orbit, but its probability of intersecting the CS is similar to that of the PSP orbit. However, the probability of Orb$_5$ passing through the CS at large angles is considered not low as $\alpha$ is large. The Orb$_6$ has the highest intersection probability and is the orbit that is most likely to cross the CS at medium or high angles among all the orbits.

\section{Probability for spacecraft to traverse CS} \label{sec:Result traverse}

In previous sections, we calculated the probability of a heliocentric orbit intersecting a CS. In this section, we investigate the probability of the spacecraft itself crossing a CS. Apparently, the spacecraft can only traverse a CS if its orbit is capable of intersecting the CS. In addition to the parameters discussed earlier, the probability of spacecraft crossing the CS is constrained by the moment $t$ when the eruption starts, the CME velocity $v_c$, and the spacecraft velocity $v_s$. In general, we assume that the time $t$ is independent of the other parameters. Therefore, the joint PDF $f(\theta_s,\phi_s,\gamma,l,t)$ can be expressed as on the basis of Equation (\ref{Density_tpgl}):
\begin{eqnarray}
f(\theta_s,\phi_s,\gamma,l,t)=f(\theta_s,\gamma)f(\phi_s)f(l)f(t) \label{Density_tpglt}.
\end{eqnarray}
It is also reasonable to assume a constant rate of the eruption within a given time interval:
\begin{eqnarray}
f(t)&=&\frac{1}{T_0}, t \in(0, T_0)\label{Density_t},
\end{eqnarray}
where, $T_{0}=1$~year. In fact, $T_{0}$ could be any value. Here, we mainly study the number of times the spacecraft may traverse the CS during one year, so we set $T_{0}=1$~year.

Repeat the steps of deriving Equations (\ref{Density_tpgl}) through (\ref{P condition}), we rewrite Equation (\ref{P define3}) for $P^{tra}$ as:
\begin{eqnarray}
P^{tra}&=&\int f(l)P^{tra}_l{\rm d}l\label{P tota2},
\end{eqnarray}
where $P^{tra}_l=P\left[(\theta_s, \phi_s, \gamma, l, t)\in \Omega_{1,2,3}~|~l\right]$ is the conditional probability of spacecraft traversing a CS with length $l$. This probability is calculated in the way:
\begin{eqnarray}
P^{tra}_l&=&\int_{\Omega_{1,2,3}^l}f(\theta_s,\gamma)f(\phi_s)f(t) {\rm d}\theta_s {\rm d}\phi_s {\rm d}\gamma{\rm d}t\label{P condition2}.
\end{eqnarray}
Because samples that could be collected here are discrete individual events, the integral in (\ref{P tota2}) could be simplified into a finite summation:
\begin{eqnarray}
P^{tra}&=&\sum_{i=1}^{N}P(l_i-\frac{\Delta l}{2}<l<l_i+\frac{\Delta l}{2})P^{tra}_{l_i}\label{P tota2.2},
\end{eqnarray}
where $N$ is the total number of samples, $l_{i}$ is the CS length of the $i$th sample, $P(l)=f(l)\Delta l$, and $P(l_{i}-\Delta l /2 < l < l_{i}+\Delta l /2)$ is the total probability of the occurrence of the CS with a length in the range of $\l_{i}\pm \Delta l/2$.

According to \cite{2016SoPh..291.3725W}, the average ratio of the speed of the CMEs, $v_c$, to the speed of the associating CS increase in length, $v$, is 2.2. Assuming a constant growth rate of the CS for simplicity, $l$ is thus related to $v_{c}$ and the life-time of CS, $\tau_{0}$, such as:
\begin{eqnarray}
l(v_c)&=&\frac{\tau_0}{2.2}v_{c}\label{l_vc}.
\end{eqnarray}
Then the probability $P(l_{min} \leq l \leq l_{max})$ of the occurrence of a CS within a certain range of the length is related to the probability $P(v_{min} \leq v_c \leq v_{max})$ within a certain growth rate range:
\begin{eqnarray}
P(l_i-\frac{\Delta l}{2}<l<l_i+\frac{\Delta l}{2})&=&P(v_{ci}-\frac{\Delta v_{c}}{2}<v_c<v_{ci}+\frac{\Delta v_{c}}{2})\label{l to v},
\end{eqnarray}
where $l_i=\tau_0 v_{ci}/2.2$, $\Delta l=\tau_0 \Delta v_{c}/2.2$. Furthermore, substituting Equation (\ref{l to v}) into (\ref{P tota2.2}) gives:
\begin{eqnarray}
P^{tra}&=&\sum_{i=1}^{N}P(v_{ci}-\frac{\Delta v_{c}}{2}<v_c<v_{ci}+\frac{\Delta v_{c}}{2})P^{tra}_{l(v_{ci})}\label{P tota2.3}.
\end{eqnarray}
For extremely slow CMEs, the trailing CS is most likely to totally dissipate before they encounter the orbit. On the other hand, the probability of the occurrence of extremely fast CMEs are too low to be traversed. Therefore, as an approximation, we only consider that the CME velocity ranges from $100$~km~s$^{-1}$ to 1100~km~s$^{-1}$. For convenience, we divide the velocity range into 11 intervals. Therefore, $N=11$, $\Delta v_c=100$~km~s$^{-1}$, and $v_{ci}=100i$~km~s$^{-1}$, $i=1$, ..., $N$. Finally, Equation (\ref{P tota2.3}) becomes:
\begin{eqnarray}
P^{tra}&\approx &P(v_c<150)P^{tra}_{l(v_{c1})}+ \sum_{i=2}^{10}P(v_{ci}-50<v_c<v_{ci}+50)P^{tra}_{l(v_{ci})}+P(v_c>1050)P^{tra}_{l(v_{c11})}\label{P tota2.4}.
\end{eqnarray}

\begin{figure}[ht]
\begin{center}
\includegraphics[width=0.9\textwidth]{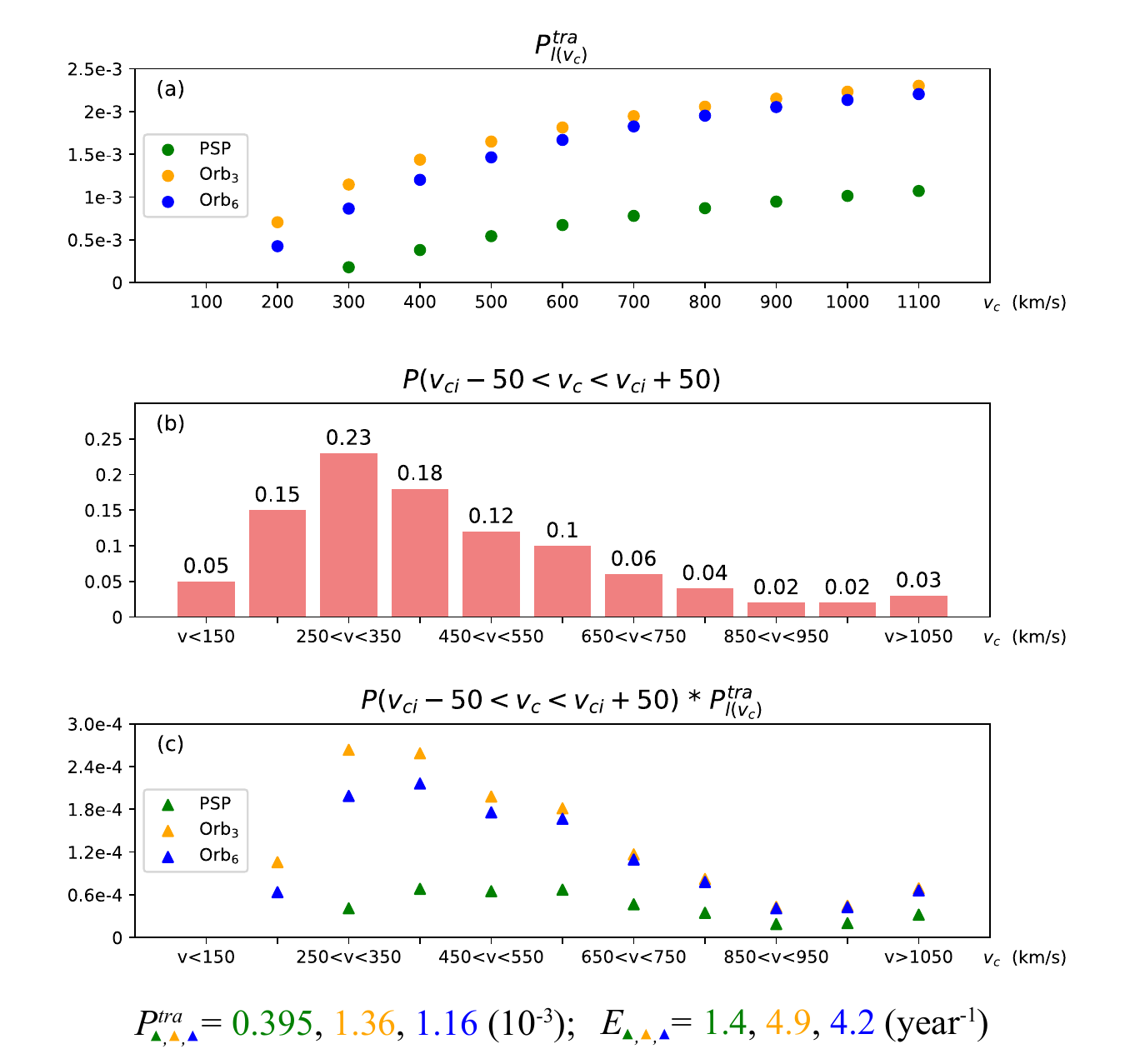}
\end{center}
\caption{(a) The conditional probability for a CS to be traversed by the spacecraft, given that the CS is generated by a CME with speed $v_c$. Here, the green, orange and blue scatter plots are results of different orbits: PSP, Orb$_3$ and Orb$_6$, respectively. (b) The probability distributions of the speed of CMEs \citep[CDAW data from][]{2019SSRv..215...39L}. (c) The probability for a CS that is produced by a CME with various velocities to be traversed, which is the product of the corresponding terms in panels (a) and (b). Unlike panel (a), panel (c) includes the impact of CME velocities on estimating the probability. At the bottom of panel (c), we present the probability for the spacecraft in different orbits to traverse a CS, $P^{tra}$ (see Equation \ref{P tota2.4}), and the expected numbers for the spacecraft to traverse the CSs in one year in the solar maximum (see Equation \ref{E}).
\label{fig:probability_diff_vmces}}
\end{figure}

Combining Equations (\ref{P condition2}) and (\ref{l_vc}), we calculate the conditional probability $P^{tra}_{l(v_{ci})}$ (see Equation \ref{P tota2.3}) of detecting a CS trailing a CME of a given speed $v_{ci}$. As mentioned before, we set $\tau_0=18$~hrs as the lower limit of the CS lifetime. We compare three different orbits, namely Orb$_1$ (PSP), Orb$_3$, and Orb$_6$, as listed in Table \ref{tab:cases_orb}, and present the results in Figure \ref{fig:probability_diff_vmces}a. The green, orange, and blue points give the results for PSP, Orb$_3$, and Orb$_6$, respectively. We notice that for $v_c<300$~km~s$^{-1}$, PSP cannot detect the CS behind the associated CME, as the CS produced by a slow CME cannot reach the PSP orbit within its lifetime, and thus does not meet Criterion 3 (see Equation \ref{Criterion3}). With increasing CME speed, the probability of detecting the associated CS also increases. \cite{2016SoPh..291.3725W} statistically analyzed the speed of 40 CMEs and the associated CSs during the solar maximum, obtaining an average speed of CME of 705~km~s$^{-1}$. We also noticed that some slow CMEs at speed as low as 100~km~s$^{-1}$ could even produce CS (e.g., see also \citealt{2002ApJ...575.1116C}).

To estimate $P^{tra}$ in a more realistic scenario, we need to consider the weight contributed by the number of CMEs with different speeds. Therefore, we plot the probability distribution of the CME occurrence, $P(v_{ci}-\Delta v_{c}/2<v_c<v_{ci}+\Delta v_{c}/2)$ (see Equation \ref{P tota2.3}), versus CME speeds in Figure \ref{fig:probability_diff_vmces}b according to \cite{2019SSRv..215...39L}. Combining Figures \ref{fig:probability_diff_vmces}a and \ref{fig:probability_diff_vmces}b, we obtain probabilities for a CS behind CME of various velocities to be the DCS (see Figure \ref{fig:probability_diff_vmces}c). Results in Figure \ref{fig:probability_diff_vmces}c are equivalent to those in Figure \ref{fig:probability_diff_vmces}a with the difference in CME speeds as the weight being included in calculations by taking the product of the corresponding values given in Figures \ref{fig:probability_diff_vmces}a and \ref{fig:probability_diff_vmces}b. Although a fast CME is more likely to generate a DCS, fast CMEs are usually less than slow ones. Therefore, in reality, the CS produced by a relatively slow CME could be detected more easily. Specifically, speeds of most CMEs are between $250$ and $550$~km~s$^{-1}$, and thus they have the highest probability of producing DCSs.

Comparing probabilities of spacecraft in different orbits traversing CSs, we find that the relatively high detection probability for a small orbit is mainly due to the advantage of detecting CSs produced by slow CMEs. Moreover, although the probability of Orb$_3$ intersecting the CS is almost the same as that of Orb$_6$, considering the motion of the spacecraft and the extension of the CS, a spacecraft in Orb$_3$ has a higher probability of traversing. Based on the collected data, we use Equation (\ref{P tota2.4}) to calculate probabilities of a spacecraft in Orb$_{1}$, Orb$_{3}$, and Orb$_{6}$ crossing a CS produced by a random solar eruption, which is equivalent to summing up each term shown in Figure \ref{fig:probability_diff_vmces}c, and obtain $P^{tra}=3.95\times10^{-4}$, $1.36\times10^{-3}$ and $1.16\times10^{-3}$, respectively (see Figure \ref{fig:probability_diff_vmces}). On the basis of these results, we are able to further estimate the expected number of the spacecraft traversing the CS in a given year.

\cite{2019SSRv..215...39L} performed a statistical analysis of the rate of the CME occurrence in solar cycles 23 and 24. The data was categorized into four groups based on different detection methods, namely ARTEMIS, SEEDS, CACTus, and CDAW. They found that the rate of CME occurrence in the first two categories is about 400/month, and that in the third and the fourth categories is about 200/month. As an approximation, we assume that 10 CMEs occur everyday in the solar maximum, so we have 3650 CMEs per year. As we mentioned earlier that the occurrence of a CS is the occurrence of a random event $\left\{\theta_s,\phi_s,\gamma,l,t \right\}$. Therefore, calculating the probability of detecting a random CS is similar to the process of randomly sampling points in a five-dimensional parameter space, in which the vector constituted by these parameters determines an individual point, and the CS is judged detectable when this point is located within domain $\Omega_{1,2,3}$.

Equation (\ref{P define3}) gives $P^{tra}$, and also determines that a corresponding point in the five-dimensional space is located in $\Omega_{1,2,3}$. The question of interest here is how many DCSs could be produced by CMEs every year? In other words, how many randomly sampled points among 3650 ones would fall within the region $\Omega_{1,2,3}$? Given that each CME event is independent of the other events, each sampling process is as an independent experiment with only two outcomes: success (the spacecraft traverses the CS) and failure (the spacecraft does not traverse the CS). The probability of success is $P^{tra}$ for each event, and the probability of failure is $1-P^{tra}$. Denote $M$ the number of successful experiments among total $n=3650$ experiments. Under the assumption of independence, $M$ follows a binomial distribution, $M\sim B(n,P^{tra})$:
\begin{eqnarray}
P(M=k)={\rm C}_n^k(P^{tra})^k(1-P^{tra})^{n-k},~k=0,1,...,n\label{Binomial distribution}.
\end{eqnarray}
The expected value of $M$ is then:
\begin{eqnarray}
E(M)=\sum_{k=0}^{n}kP(M=k)=\sum_{k=0}^{n}k{\rm C}_n^k(P^{tra})^k(1-P^{tra})^{n-k}=nP^{tra}.\label{E}
\end{eqnarray}
Combining Equations (\ref{P tota2.4}) and (\ref{E}) gives the expected number of the spacecraft traversing the CS in different orbits per year in the solar maximum. Multiplying the $P^{tra}$ in different orbits calculated earlier by $n$, we obtain the expected number of the spacecraft in Orb$_{1}$, and Orb$_{2}$, and Orb$_{3}$ traversing the CS per year $E(M)=1.4$, $4.9$, and $4.2$, respectively (see Figure \ref{fig:probability_diff_vmces}). Therefore, the probability of PSP traversing a CME-flare CS is not high because: 1. the inclination of the PSP orbit is small, and the orbit of large inclination would allow high probability of traversing; 2. the perihelion of PSP orbit is still far away from the Sun.

An intriguing question is whether any spacecraft has ever detected a CME-flare CS yet. PSP, Solar Orbiter, and Bepi-Colombo \citep[][]{2010P&SS...58....2B} might be the three candidates in orbit. Nour E. Raouafi (private communications) mentioned that a traversal was very likely to occur on September 5, 2022 because a major eruption produced a fast CME that swept the PSP first, and then the PSP traversed the CS behind the CME at a fairly large angle, $\sigma$. \cite{Romeo_2023} reported that reversal of the radial component of the magnetic field was detected by PSP, and proposed that the reversing might result from the traverse of a CME-flare CS as described by \cite{2000JGR...105.2375L}. As of today, the Solar Orbiter has not been reported to cross a candidate of the CME-flare CS yet, it is probably too far from the Sun to encounter anything like a CS (Angelos Vourlidas, private communications). As for the Bepi-Colombo, it is a mission of orbiting Mercury with the perihelion of 65.6~$R_{\odot}$ and aphelion of 99.8~$R_{\odot}$, and after comparing with the Solar Orbiter with perihelion of 59.8~$R_{\odot}$, we might not be able to expect a traverse of a CME-flare CS by Bepi-Colombo.

\section{Conclusions} \label{sec:Discussion}

Although traversing the large-scale CME-flare CS was not initially among the scientific goals of the PSP mission, the probability exists that PSP traverses CSs and provides important and essential information on the large-scale CSs and the magnetic reconnection processes therein. Due to the randomness of solar eruptions in both space and time, not all orbits are expected to allow the spacecraft to traverse CSs, for example, our calculations indicate that the PSP orbit is not the optimal one for crossing CSs. Based on the Lin-Forbes model and existing observations, we utilized the GCS model for CME/ICME reconstruction developed by \cite{2009SoPh..256..111T} and employed a method to calculate the probability of PSP or similar spacecraft traversing the CS generated by a solar eruption. We simplified the CS as a triangle-shaped plate, established a quantitative relationship between the relevant parameters of the DCSs and the orbits, and then estimated the probability of a PSP-like probe crossing the CS on given orbits.

Three criteria were established to check whether a CME-flare CS could be traversed by a spacecraft in a given orbit. The first criterion checks whether the orbit of spacecraft could cross the CS, namely whether at least two points exist on the CS such that these two points are located either side of the orbital plane. Criterion 2 requires that at least one point on the CS-orbit intersection is located outside the ellipse of the orbit, and criterion 3 determines the condition under which a spacecraft itself crosses the CS. A spacecraft could traverse a CME-flare CS successfully only if these three criteria are satisfied simultaneously.

Our results show that the CS could be traversed by the spacecraft easily if the corresponding eruption propagates roughly along the plane of the spacecraft orbit, i.e., the symmetry axis of CS and CME almost lies in the orbital plane. In addition, because of the finite length and lifetime of the CS, as well as the finite speed at which the spacecraft moves, the traverse is more likely to happen if the eruption that produces the CS occurs in the region near the perihelion.

On the basis of the existing cases of the solar eruption and the distribution of source regions of these eruptions (\citealt{2015ApJS..221...33H}), we investigated carefully various possible relative positions of the CME-flare CS produced in these events and a given orbit of the spacecraft with the purpose to detect CSs. We found that an orbit of inclination, $\alpha > 10^{\circ}$, to the ecliptic plane would help enhance the probability of the spacecraft traversing CSs. Considering the fact that traversing the CS orthogonally is very hard, if not impossible, we studied the probability for the satellite to traverse the CS occurring at the angle, $\sigma$, of medium values, say $30^{\circ}<\sigma<60^{\circ}$, and obtained a probability around 0.1\% as $\alpha > 30^{\circ}$. In the solar maximum, the total number of traversing the CS by a spacecraft on such an orbit is about 4/year. The probability for PSP to traverse a CS is around 0.04\%, and the expected number of traversing the CS is about 1.4/year.

\vspace{2em}
\noindent
Authors are appreciating the referee for the valuable comments and suggestions that helped improve this work. We gratefully acknowledge constructive comments and suggestions given by Weiqun Gan, Terry G. Forbes and John C. Raymond. This work was supported by National Key R\&D Program of China No. 2022YFF0503804, the NSFC grant 11933009, 12273107 and U2031141, grants associated with the Yunling Scholar Project of the Yunnan Province, the Yunnan Province Scientist Workshop of Solar Physics, and the Applied Basic Research of Yunnan Province 2019FB005. The numerical computation in this paper was carried out on the computing facilities of the Computational Solar Physics Laboratory of Yunnan Observatories (CoSPLYO).

\appendix

\section{Formulations for Deducing coordinates of $C_1$ and $C_2$}\label{sec:appA}

For a CS described by the parameters of ($l$, $\delta$, $\phi_{s}$, $\theta_{s}$, $\gamma$), the calculation of the coordinates of $C_1$ and $C_2$ in the $XYZ$-system includes two steps. First, we consider a referential case with $\theta_{s}=\gamma=0$. So the coordinates of $C_{1a}$ and $C_{2a}$, which are the counterparts of $C_{1}$ and $C_{2}$ in the referential case, can be easily obtained. Second, we transform the CS configuration from the referential case into the true case by performing three rotations, which eventually gives the coordinates of $C_1$ and $C_2$. The relevant transformations are illustrated in Figure \ref{fig:Rotation} with detailed explanations given below.

\begin{figure}[ht]
\begin{center}
\includegraphics[width=1.0\textwidth]{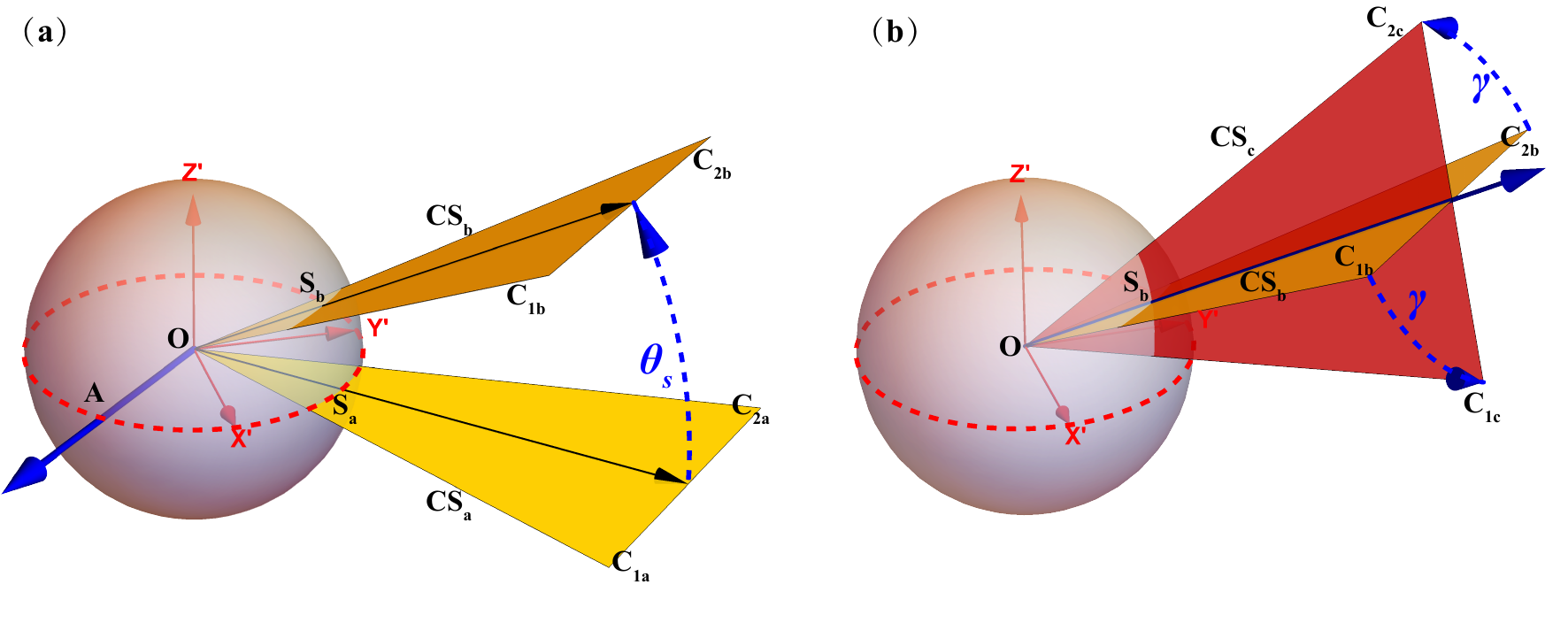}
\end{center}
\caption{A sketch of the rotating operations that correspond to rotating matrices $\mathcal{M}(\bf{OA},\theta_s)$ and $\mathcal{M}(\bf{OS_b},\gamma)$, illustrating how to obtain the endpoints of a CS in $X'Y'Z'$ coordinate system by rotating the reference CS, CS$_a$.
\label{fig:Rotation}}
\end{figure}

First, we consider CS$_{a}$ as the referential CS and assume that it lies initially in the ecliptic plane (see the yellow triangle in Figure \ref{fig:Rotation}a). The symmetry axis of CS$_{a}$ extends outward along the vector $\bf{OS_{a}}$, and locations of $C_{1a}$ and $C_{2a}$, in the $X^{\prime}Y^{\prime}Z^{\prime}$ system could be easily obtained by writing down two corresponding vectors: $\bf{OC_{1a}}$$=[l\cos(\phi_{s}-\delta), l\sin(\phi_{S}-\delta), 0]$ and $\bf{OC_{2a}}$$=[l\cos(\phi_{S}+\delta), l\sin(\phi_{S}+\delta), 0]$.

We then rotate the CS$_{a}$ counterclockwise an angle $\theta_{s}\neq 0$ around $\bf{OA}$, a vector perpendicular to $\bf{OS_{a}}$ and lying in the ecliptic plane. This rotation transforms the CS$_{a}$ into CS$_{b}$ (see another yellow triangle in Figure \ref{fig:Rotation}a). The relevant parameters of CS change to $S_{b}$, $C_{1b}$, and $C_{2b}$ accordingly. The coordinates of $C_{1b}$ and $C_{2b}$ in $X^{\prime}Y^{\prime}Z^{\prime}$ can be expressed as follows:
\begin{eqnarray}
{\bf{OC_{1b}}}&=&\mathcal{M}(\bf{OA},\theta_s) \bf{OC_{1a}}^\mathrm{T}\label{RotateOC_{1b}},\\
{\bf{OC_{2b}}}&=&\mathcal{M}(\bf{OA},\theta_s) \bf{OC_{2a}}^\mathrm{T}\label{RotateOC_{2b}},
\end{eqnarray}
where $\mathcal{M}(\bf{OA},\theta_s)$ is the matrix for the rotation by angle $\theta_s$ around $\bf{OA}$. We further rotate CS$_b$ counterclockwise by angle $\gamma\neq 0$ around $\bf{OS_b}=(\cos{\phi_s}\cos{\theta_s},\sin{\phi_s}\cos{\theta_s},\sin{\theta_s})$, the symmetry axis of the CS, to approach to the original status of the CS described by $(l,\delta,\phi_s,\theta_s, \gamma)$. Therefore, we can express the two endpoints of CS$_c$, $C_ {1c}$ and $C_{2c}$, in the $X'Y'Z'$ coordinate system as below:
\begin{eqnarray}
{\bf{OC_{1c}}}&=&\mathcal{M}(\bf{OS_b},\gamma) \bf{OC_{1b}}^\mathrm{T}\label{RotateOC_{1c}},\\
{\bf{OC_{2c}}}&=&\mathcal{M}(\bf{OS_b},\gamma) \bf{OC_{2b}}^\mathrm{T}\label{RotateOC_{2c}}.
\end{eqnarray}
So far, we have finalized the description of any CS in the $X'Y'Z'$ system. The final step toward describing the CS morphology in the $XYZ$ system is to transform the above descriptions of CS in the $X'Y'Z'$ system into the $XYZ$ system. As shown in Figure \ref{fig:Coordinate}, two ways exist that the orbital plane ($XYZ$-system) deviates from the ecliptic plane ($X^{\prime}Y^{\prime}Z^{\prime}$-system), which means that we have two choices for the transformation in this step, rotate the CS around either $X^{\prime}$- or $Y^{\prime}$-axis clockwise by an angle of $\alpha$:
\begin{eqnarray}
{\bf{OC_1}}&=&\mathcal{M}(rot,-\alpha) \bf{OC_{1c}}^\mathrm{T}\label{RotateOC_{1}},\\
{\bf{OC_2}}&=&\mathcal{M}(rot,-\alpha) \bf{OC_{2c}}^\mathrm{T}\label{RotateOC_{2}},
\end{eqnarray}
where $rot$ means either the $X^{\prime}$- or the $Y^{\prime}$-axis. Eventually, combining Equations (\ref{RotateOC_{1b}})-(\ref{RotateOC_{2}}) gives the coordinates of $C_1$ and $C_2$ in the $XYZ$-system:
\begin{eqnarray}
{\bf{OC_1}}&=&\begin{bmatrix}
                x_1 \\
                y_1 \\
                z_1 \\

              \end{bmatrix}=\mathcal{M}(rot,-\alpha)\mathcal{M}(\bf{OS_b},\gamma)\mathcal{M}(\bf{OA},\theta_s)
\begin{bmatrix}
  l\cos{(\phi_s-\delta)} \\
  l\sin{(\phi_s-\delta)} \\
  0
\end{bmatrix}\label{RotateFinalOC_{1}},\\
{\bf{OC_2}}&=&\begin{bmatrix}
                x_2 \\
                y_2 \\
                z_2
              \end{bmatrix}=\mathcal{M}(rot,-\alpha)\mathcal{M}(\bf{OS_b},\gamma)\mathcal{M}(\bf{OA},\theta_s)
              \begin{bmatrix}
  l\cos{(\phi_s+\delta)} \\
  l\sin{(\phi_s+\delta)} \\
  0
\end{bmatrix}\label{RotateFinalOC_{2}}.
\end{eqnarray}

\section{Estimations of $\tau_{fly}$}\label{sec:appB}

\begin{figure}[ht]
\begin{center}
\includegraphics[width=0.7\textwidth]{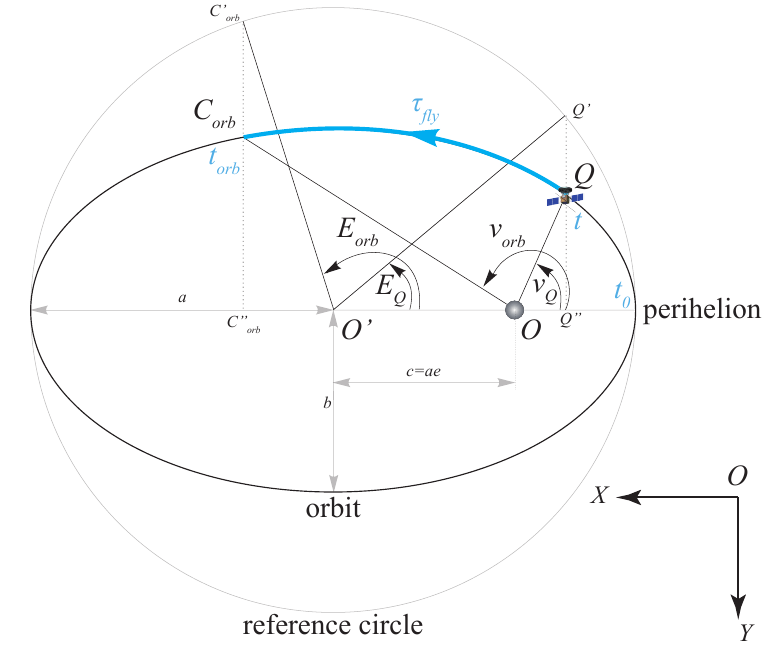}
\end{center}
\caption{True and eccentric anomalies $\nu$ and $E$. From \cite{2005mcmp.book.....B}.
\label{fig:Angle}}
\end{figure}

According to Equation (\ref{tau_fly}), evaluating $\tau_{fly}$ needs to know $t$ and $t_{orb}$, the times at which the spacecraft reaches the points $Q$ and $C_{orb}$, respectively. Therefore, a reference time is expected. We choose the time $t_0$ when spacecraft passes the perihelion as a referential point for the following reasons. Calculations for $t_{orb}$ involves evaluations of the eccentric anomaly, $E$, measured in the same direction as measuring the true anomaly, $\nu$. Figure \ref{fig:Angle} specifies the definition of $E$ and $\nu$: the Sun is located at one focus of the spacecraft orbit, $O$; the spacecraft is at point $Q$ on the orbit, which has projections $Q^{\prime}$ and $Q^{\prime\prime}$ on the reference circle of the orbit and on the major-axis of the orbit ellipse, respectively; similarly, the CS intersects the orbit at point $C_{orb}$, with the corresponding projection points $C_{orb}^{\prime}$ and $C_{orb}^{\prime\prime}$ on the reference circle and the major axis of the orbit ellipse, respectively. So the eccentric and true anomalies for point $Q$ are denoted as $E_Q=\angle OO^{\prime}Q^{\prime}$ and $\nu_Q=\angle Q^{\prime\prime}OQ$, respectively. Similarly, for point $C_{orb}$, the corresponding angles are $E_{orb}=\angle OO^{\prime}C_{orb}^{\prime}$ and $\nu_{orb}=\angle Q^{\prime\prime}OC_{orb}$. The perihelion corresponds to $E_0=0$ and $\nu_0=0$ simultaneously (see Figure \ref{fig:Angle}), and we set the time, $t_{0}=0$. This choice of reference time simplifies calculations (see \citealt{2005mcmp.book.....B} for more details).

According to \cite{2005mcmp.book.....B}, the flying time, $t_{orb}$, is
\begin{eqnarray}
t_{orb}=\sqrt{\frac{a^3}{GM_{\odot}}}(E_{orb}-e \sin{E_{orb}}) \label{Kepler equation}
\end{eqnarray}
on the basis of Kepler equation. $E_{orb}$ is related to $\nu_{orb}$ in the way of
\begin{eqnarray}
E_ {orb}=2\arctan{(\sqrt{\frac{1-e}{1+e}})}\tan{(\frac{\nu_{orb}}{2})}\label{nu2E}.
\end{eqnarray}
In the coordinate system $XYZ$ used in this work (see also Figure \ref{fig:Angle}), a phase difference, $\pi$, exists between angle $\nu_{orb}$ and the longitude $\phi$ of point $C_{orb}$:
\begin{eqnarray}
\nu_{orb}=-\pi+\phi\label{phi2nu}.
\end{eqnarray}

Given the parameters of a CS and an orbit, we can calculate $\tau_{fly}$ and examine Criterion 3 by substituting $\nu_{orb}$ in Equation (\ref{phi2nu}) into (\ref{nu2E}) for $E_{orb}$, substituting resultant $E_{orb}$ into Equation (\ref{Kepler equation}) for $t_{orb}$, and finally substituting the resultant $t_{orb}$ into Equation (\ref{tau_fly}).




\bibliography{cyh}{}
\bibliographystyle{aasjournal}


\end{CJK*}
\end{document}